\documentclass[prd,showkeys,twocolumn,amsmath,amssymb]{revtex4}
\usepackage{graphicx}
\usepackage{epstopdf}
\usepackage{subfigure}
\usepackage{bm}
\usepackage{soul} 
\usepackage[colorlinks,citecolor=blue,anchorcolor=red,menucolor=red,linkcolor=red,filecolor=red,runcolor=red,urlcolor=blue,frenchlinks=red]{hyperref}
\usepackage{ulem}
\usepackage{color}
\usepackage{multirow}

\makeatletter
\renewcommand{\@thesubfigure}{\hskip\subfiglabelskip}
\makeatother

\def\f(s){\left[(\alpha+\beta)m_c^2-\alpha\beta s\right]}

\allowdisplaybreaks[3]
\begin{document}
%=====================================================================================
%=====================================================================================
\title{QCD sum rule study on excited light meson operators}
%=====================================================================================
%=====================================================================================
%

\author{Wei-Han Tan}
\author{Wen-Ying Liu}
\author{Hong-Zhou Xi}
\author{Hua-Xing Chen}
\email{hxchen@seu.edu.cn}

\affiliation{
School of Physics, Southeast University, Nanjing 210094, China
}
\begin{abstract}
We apply the QCD sum rule method to systematically study excited light meson operators and calculate their decay constants. These operators are constructed by explicitly adding one covariant derivative to the quark-antiquark pair. In total, twelve such operators are constructed, among which ten are subjected to detailed numerical analyses. The considered quark contents include $\bar{q}q$, $\bar{q}s$, and $\bar{s}s$ ($q = u/d$), allowing the formation of various $SU(3)$ flavor nonets. For instance, our results support the interpretation that the $a_2(1320)$, $f_2(1270)$, $f_2^\prime(1525)$, and $K_2^*(1430)$ constitute a flavor nonet with quantum numbers $J^{P(C)} = 2^{+(+)}$. In addition, we predict several excited meson states, whose masses and decay constants are determined using the QCD sum rule method.
\end{abstract}
\maketitle
\pagenumbering{arabic}

\section{Introduction}
\label{sec:intro}

Within the conventional quark model, mesons are understood as bound states of a quark and an antiquark, while baryons are described as composite systems of three quarks~\cite{Gell-Mann:1964ewy,Zweig:1964ruk,Zweig:1964jf}. Over the past several decades, numerous light mesons have been observed experimentally~\cite{pdg,Belle:2009xpa,COMPASS:2014vkj,JPAC:2018zyd,COMPASS:2009xrl,E852:2001ikk,E852:1997gvf,VES:1995nsw,DeRujula:1975qlm,BES:2003iac,BESIII:2013qqz,BESIII:2019apb,SND:2023gan,SND:2021amz,Lichard:2023ngm,Achasov:2019duv,BESIII:2022riz,BESIII:2022iwi,BES:2004twe,BES:2006ssn,Belle:2008bmg,COMPASS:2020yhb,LHCb:2017swu,LHCb:2024tpv,LHCb:2013ged,BESIII:2021qfo,Brandenburg:1975gv,LHCb:2021uow,ATLAS:2024ext,BNL-E852:2000poa}, prompting extensive theoretical investigations. These studies have employed a wide range of approaches, including the quark model~\cite{Godfrey:1985xj,DeRujula:1975qlm}, the MIT bag model~\cite{Chodos:1974je,Theberge:1980ye,Freedman:1976ub,Friedberg:1977xf,Johnson:1975zp}, the flux-tube model~\cite{Isgur:1984bm,Casher:1978wy,Barnes:1995hc}, the AdS/QCD model~\cite{deTeramond:2005su,Brodsky:2014yha,Roberts:2021nhw,Li:2013oda}, Lattice QCD~\cite{Kogut:1982ds,Asakawa:2000tr,Alford:2000mm,Nakahara:1999vy,Abada:1991mt,McNeile:2006qy,Edwards:2011jj,Dudek:2010wm,HadronSpectrum:2012gic,Dudek:2011tt,Dudek:2016cru,HadronSpectrum:2009krc,Shultz:2015pfa,Woss:2019hse,Bulava:2011np,Morningstar:2013bda,Dudek:2024roh,Chen:2022isv,Briceno:2017qmb}, and QCD sum rules~\cite{Shifman:1978bx,Kwiecinski:1984rn,Colangelo:2000dp,Azizi:2014maa,Huang:1998zj,Latorre:1985tg,Yang:2007cc,Tan:2024grd,Aliev:2009nn,Xin:2022qnv,Wang:2019nln,Wang:2015uha,Chen:2007xr,Su:2022eun,Yu:2019kcr,Yu:2021ggd,Azizi:2017izn,Agaev:2017lmc,Wang:2016hoi,Wang:2014yza,Aliev:2015sea,Sundu:2012zz}, among others.

When applying the QCD sum rule method to investigate meson states, one requires suitable meson operators, each composed of a quark field and an antiquark field combined with an appropriate Lorentz structure $\Gamma$:
\begin{equation}
J = \bar q_a \Gamma q_a \, ,
\end{equation}
where $a$ denotes the color index. Without the inclusion of covariant derivatives, there exist five independent local operators:
\begin{eqnarray}
\nonumber &\bar q_a  \gamma_5 q_a \, , \, \bar q_a  q_a \, , \, &
\\ & \bar q_a  \gamma_\mu q_a \, , \, \bar q_a  \gamma_\mu \gamma_5 q_a \, , \, &
\\ \nonumber & \bar q_a  \sigma_{\mu\nu} q_a \, , &
\end{eqnarray} 
Additionally, the operator 
\begin{equation}
\bar q_a \sigma_{\mu\nu} \gamma_5 q_a\, ,
\end{equation} 
is not independent, as it can be related to $\bar q_a \sigma_{\mu\nu} q_a$ via the identity $\sigma_{\mu\nu} \gamma_5 = {i/2} \times \epsilon_{\mu\nu\alpha\beta} \sigma^{\alpha\beta}$. These ``ground-state'' meson operators have been extensively studied in the literature. In contrast, ``excited'' meson operators involving covariant derivatives have received relatively little attention and merit further exploration.

In this paper we shall systematically investigate excited light meson operators using the QCD sum rule method. Each operator is constructed from a quark field, an antiquark field, and a covariant derivative, combined with an appropriate Lorentz structure $\Gamma$. Since the excited states are expected to correspond to internal orbital angular momentum excitations rather than collective excitations of the meson, we have chosen to place the covariant derivatives between the quark and antiquark fields in our operator construction:
\begin{equation}
J = \bar q_a {\overset{\leftrightarrow}{D}}_\alpha \Gamma q_a \, ,
\label{def:struct}
\end{equation}
 where the covariant derivative is defined as
\begin{equation}
\big[ X {\overset{\leftrightarrow}{D}}_\alpha Y \big] = X (D_\alpha Y) - (D_\alpha X) Y \, ,
\end{equation}
with
\begin{equation}
D_\alpha q_a = \partial_\alpha q_a + i g_s A_\alpha q_a  = \partial_\alpha q_a + i g_s A^n_\alpha {\lambda^n_{ab}\over2 } q_b \, .
\end{equation}
These operators can couple to meson states through
\begin{equation}
\langle 0 | J | X \rangle = f_X \, ,
\end{equation}
where $f_X$ denotes the decay constant. In total, we shall construct twelve excited light meson operators and carry out detailed analyses for ten of them. By specifying the quark content as $\bar{q}q$, $\bar{q}s$, or $\bar{s}s$ ($q = u/d$), we shall apply the QCD sum rule method to compute their masses and decay constants.

This paper is organized as follows. In Sec.~\ref{sec:current} we present a systematic construction of excited light meson operators. Building on this foundation, we perform QCD sum rule analyses using a subset of these operators in Sec.~\ref{sec:sumrule}. Finally, in Sec.~\ref{sec:summary} we summarize our findings and provide a detailed discussion of the results.

\section{Phenomenological analyses}
\label{sec:current}

\begin{figure*}[hbt]
\begin{center}
\includegraphics[width=0.86\textwidth]{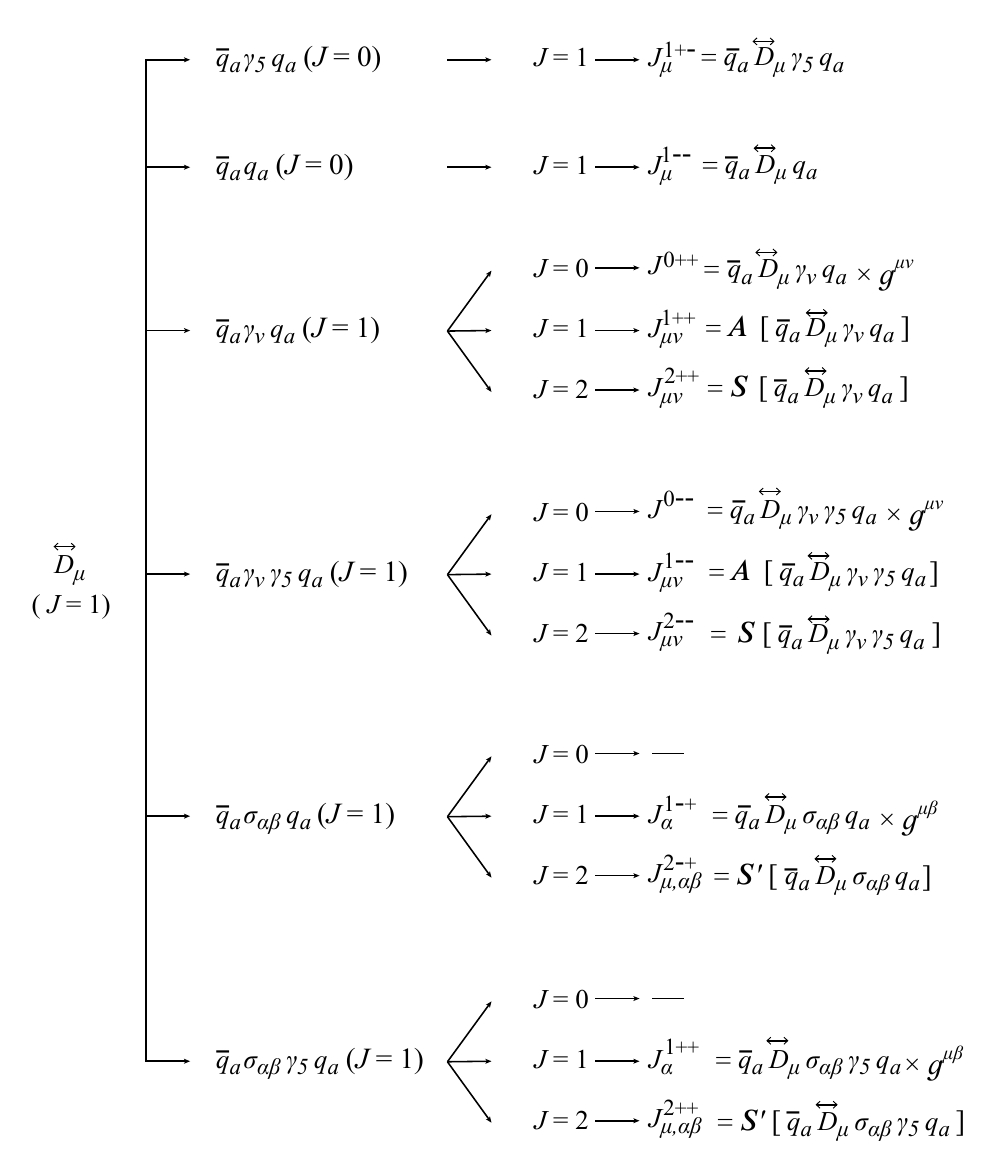}
\caption{Categorization of excited light meson operators:  Each operator is constructed from a quark field, an antiquark field, and a covariant derivative, combined with an appropriate Lorentz structure \( \Gamma \). The symbol \( \mathcal{A}[\cdots] \) denotes anti-symmetrization over the index set \( \{\mu\nu\} \),  \( \mathcal{S}[\cdots] \) represents symmetrization and trace subtraction over \( \{\mu\nu\} \), and \( \mathcal{S^\prime}[\cdots] \) indicates symmetrization and trace subtraction over the set \( \{\mu\alpha\} \).}
\label{fig:category}
\end{center}
\end{figure*}

In this section we systematically construct excited light meson operators by combining one quark field, one antiquark field, one covariant derivative, and appropriate Lorentz structures, as shown in Eq.~(\ref{def:struct}). For example, there are several ways to combine the covariant derivative ${\overset{\leftrightarrow}{D}}_\alpha$ with the tensor operator $\bar q_a \sigma_{\mu\nu} q_a$:
\begin{itemize}

\item One approach is to leave all Lorentz indices uncontracted:
\begin{equation}
J_{\mu,\alpha\beta} = \bar q_a {\overset{\leftrightarrow}{D}}_\mu \sigma_{\alpha\beta} q_a \, .
\label{eq:example}
\end{equation}
This operator generally contains spin-1 and spin-2 components and may also include a spin-0 component. It does not contain a spin-3 component, as the indices $\alpha$ and $\beta$ are antisymmetric. The spin-2 component can be projected using
\begin{equation}
J_{\mu,\alpha\beta}^{2{-+}} = \mathcal{S^\prime}[\bar q_a {\overset{\leftrightarrow}{D}}_\mu \sigma_{\alpha\beta} q_a ] \, ,
\end{equation}
where \( \mathcal{S^\prime}[\cdots] \) denotes symmetrization and trace subtraction over the set \( \{\mu,\alpha\} \), implemented via the projection operator
\begin{eqnarray}
&& \Gamma_{\alpha^\prime\mu^\prime\nu^\prime;\alpha\mu\nu}
\\ \nonumber &=& g_{\alpha^\prime \alpha} g_{\mu^\prime \mu} g_{\nu^\prime \nu} 
+ g_{\mu^\prime \alpha} g_{\alpha^\prime \mu} g_{\nu^\prime \nu} 
- \frac{1}{2} g_{\alpha^\prime \mu^\prime} g_{\alpha \mu} g_{\nu^\prime \nu} \, .
\end{eqnarray}

\item The spin-1 component of Eq.~(\ref{eq:example}) can be more conveniently obtained by index contraction:
\begin{equation}
J_{\alpha}^{1{-+}} = \bar q_a {\overset{\leftrightarrow}{D}}_\mu \sigma_{\alpha\beta} q_a \times g^{\mu\beta} \, .
\end{equation}

\item Eq.~(\ref{eq:example}) may also contain a spin-0 component, though we currently do not know how to isolate it in a relativistically covariant form.

\end{itemize}

As summarized in Fig.~\ref{fig:category}, a total of twelve such operators are constructed and analyzed in this work, denoted as \( J^{J{PC}}_{\cdots} \), where \( J{PC} \) represents the quantum numbers of each operator: \( J \) denotes the total angular momentum, \( P \) the parity, and \( C \) the charge-conjugation parity. The operators are defined as follows:
\begin{eqnarray}
J_{\mu}^{1{+-}} &=& \bar q_a {\overset{\leftrightarrow}{D}}_\mu\gamma_5  q_a \, ,
\label{def:Ja1pm}
\\
J_{\mu}^{1{--}} &=& \bar q_a {\overset{\leftrightarrow}{D}}_\mu  q_a \, ,
\label{def:Ja1mm}
\\
J^{0{++}} &=&  \bar q_a {\overset{\leftrightarrow}{D}}_\mu\gamma_{\nu}  q_a \times g^{\mu\nu}\, ,
\label{def:J0pp}
\\
J_{\mu\nu}^{1{++}} &=& \mathcal{A}[\bar q_a {\overset{\leftrightarrow}{D}}_\mu\gamma_{\nu}  q_a ]\, ,
\label{def:Jam1pp}
\\
J_{\mu\nu}^{2{++}} &=&  \mathcal{S}[\bar q_a {\overset{\leftrightarrow}{D}}_\mu\gamma_{\nu}  q_a ]\, ,
\label{def:Jam2pp}
\\
J^{0{--}} &=&  \bar q_a {\overset{\leftrightarrow}{D}}_\mu\gamma_{\nu}\gamma_5 q_a \times g^{\mu\nu}\, ,
\label{def:J0mm}
\\
J_{\mu\nu}^{1{--}} &=&  \mathcal{A}[\bar q_a {\overset{\leftrightarrow}{D}}_\mu\gamma_{\nu}\gamma_5  q_a ]\, ,
\label{def:Jam1mm}
\\
J_{\mu\nu}^{2{--}} &=&  \mathcal{S}[\bar q_a {\overset{\leftrightarrow}{D}}_\mu\gamma_{\nu}\gamma_5  q_a ]\, ,
\label{def:Jam2mm}
\\
J_{\alpha}^{1{-+}} &=&  \bar q_a {\overset{\leftrightarrow}{D}}_\mu \sigma_{\alpha\beta}  q_a \times g^{\mu\beta} \, ,
\label{def:Jm1mp}
\\
J_{\mu,\alpha\beta}^{2{-+}} &=& \mathcal{S^\prime}[\bar q_a {\overset{\leftrightarrow}{D}}_\mu \sigma_{\alpha\beta}  q_a ]\, ,
\label{def:A2mp}
\\
J_{\alpha}^{1{++}} &=&  \bar q_a {\overset{\leftrightarrow}{D}}_\mu \sigma_{\alpha\beta} \gamma_5  q_a  \times g^{\mu\beta} \, ,
\label{def:Jm1pp}
\\
J_{\mu,\alpha\beta}^{2{++}} &=& \mathcal{S^\prime}[\bar q_a {\overset{\leftrightarrow}{D}}_\mu \sigma_{\alpha\beta} \gamma_5 q_a ]\, .
\label{def:B1mm}
\end{eqnarray}
The derivations of their quantum numbers will be presented in detail in Sec.~\ref{sec:summary}. In the above expressions, the symbol $\mathcal{A}[\cdots]$ denotes anti-symmetrization over the index set $\{\mu\nu\}$, which is implemented by applying the projection operator
\begin{eqnarray}
\Gamma_{\mu^\prime\nu^\prime;\mu\nu} &=& g_{\mu^\prime \mu} g_{\nu^\prime \nu} - g_{\nu^\prime \mu} g_{\mu^\prime \nu} \, .
\end{eqnarray}
The symbol $\mathcal{S}[\cdots]$ represents symmetrization and trace subtraction over $\{\mu\nu\}$, achieved using the projection operator
\begin{eqnarray}
 \Gamma_{\mu^\prime\nu^\prime;\mu\nu}
 = g_{\mu^\prime \mu} g_{\nu^\prime \nu}  + g_{\nu^\prime \mu} g_{\mu^\prime \nu}  - \frac{1}{2} g_{\mu^\prime \nu^\prime} g_{\mu \nu}  \, .
\end{eqnarray}

Due to the substantial computational complexity associated with the operators $J_{\mu\alpha,\beta}^{2{-+}}$ and $J_{\mu\alpha,\beta}^{2{++}}$, we have not carried out a QCD sum rule analysis for these specific cases. Instead, we refer the reader to Ref.~\cite{Wang:2023whb} for a detailed discussion.  In addition, the operator given in Eq.~(\ref{def:J0pp}) can be transformed through the equations of motion for the quark fields as:
\begin{eqnarray}
\bar{q}_a \overset{\leftrightarrow}{D}_\mu \gamma_{\nu} q_a \times g^{\mu\nu} 
&=& -2im_q \bar{q}_a q_a \, .
\end{eqnarray}
Similarly, the operator given in Eq.~(\ref{def:J0mm}) can be transformed as:
\begin{eqnarray}
\bar{q}_a \overset{\leftrightarrow}{D}_\mu \gamma_{\nu}\gamma_5 q_a \times g^{\mu\nu} 
&=& 0 \, .
\end{eqnarray}
Therefore, we do not use them to perform QCD sum rule analysis.

Since each operator can, in general, couple to multiple states, we analyze the Lorentz structures of the excited light meson operators listed above individually in the following:
\begin{itemize}

\item The operators $J^{0{++}}$ and $J^{0{--}}$ contain only the $J^{PC} = 0^{++}$ and $0^{--}$ components, respectively. Accordingly, they couple exclusively to the corresponding spin-0 meson states via
\begin{eqnarray}
\langle 0 | J^{0{++}} | X_{0{++}} \rangle &=& f_{0{++}} \, ,
\\ \langle 0 | J^{0{--}} | X_{0{--}} \rangle &=& f_{0{--}} \, ,
\end{eqnarray}
where $f_{0{++}}$ and $f_{0{--}}$ denote the decay constants.

\item The operators \( J_{\mu\nu}^{2{++}} \) and \( J_{\mu\nu}^{2{--}} \) contain only the \( J^{PC} = 2^{++} \) and \( 2^{--} \) components, respectively. Accordingly, they couple exclusively to the corresponding spin-2 meson states via
\begin{eqnarray}
\langle 0 | J_{\mu\nu}^{2{++}} | X_{2{++}} \rangle &=& \epsilon_{\mu\nu} f_{2{++}} \, ,
\label{eq:coupling7} \\
\langle 0 | J_{\mu\nu}^{2{--}} | X_{2{--}} \rangle &=& \epsilon_{\mu\nu} f_{2{--}} \, ,
\label{eq:coupling8}
\end{eqnarray}
where \( f_{2{++}} \) and \( f_{2{--}} \) are the decay constants, and \( \epsilon_{\mu\nu} \) is the polarization tensor with two symmetric Lorentz indices \( \mu\nu \).

\item The operator \( J_{\mu\nu}^{1{--}} \), with two antisymmetric Lorentz indices \( \{\mu\nu\} \), contains both \( J^{PC} = 1^{--} \) and \( 1^{+-} \) components. Therefore, it can couple to meson states with these quantum numbers through
\begin{eqnarray}
\langle 0 | J_{\mu\nu}^{1{--}} | X_{1{--}} \rangle &=& i f_{1{--}} \epsilon_{\mu\nu \alpha \beta} \epsilon^\alpha q^\beta \, ,
\label{eq:coupling3} \\
\langle 0 | J_{\mu\nu}^{1{--}} | X_{1{+-}} \rangle &=& i f_{1{+-}} (q_\mu \epsilon_\nu - q_\nu \epsilon_\mu) \, ,
\label{eq:coupling4}
\end{eqnarray}
where \( f_{1{--}} \) and \( f_{1{+-}} \) are the decay constants, \( q_\mu \) is the four-momentum, \( \epsilon_\mu \) is the polarization vector, and \( \epsilon_{\mu\nu \alpha \beta} \) denotes the Levi-Civita symbol in four dimensions, which is the fully antisymmetric tensor. Since the Lorentz structures in Eq.~(\ref{eq:coupling3}) and Eq.~(\ref{eq:coupling4}) are totally different, we can clearly separate the contributions of the two states $X_{1{--}}$ and $ X_{1{+-}}$ at the hadron level. Specifically, we can isolate $X_{1{--}}$ by considering the two-point correlation function
\begin{eqnarray}
&& \langle 0 | J_{\mu\nu}^{1{--}} | X_{1{--}} \rangle \langle X_{1{--}} | J_{\mu^\prime\nu^\prime}^{1{--}\dagger} | 0 \rangle
\\ \nonumber &=& f_{1{--}}^2 ~ \epsilon_{\mu\nu\alpha\beta} \epsilon^\alpha q^\beta ~ \epsilon_{\mu^\prime\nu^\prime\alpha^\prime\beta^\prime} \epsilon^{*\alpha^\prime} q^{\beta^\prime}
\\ \nonumber &=& - f_{1{--}}^2 ~[ q^2  \left( g_{\mu \mu^\prime} g_{\nu \nu^\prime} - g_{\mu \nu^\prime} g_{\nu \mu^\prime} \right)
- g_{\mu \nu^\prime}q_\nu q_{\mu^\prime}
\\ \nonumber&& - g_{\nu \mu^\prime}q_\mu q_{\mu^\prime} + g_{\mu \mu^\prime}q_\nu q_{\nu^\prime} + g_{\nu \nu^\prime}q_\mu q_{\mu^\prime}]  \, ,
\label{eq:coupling5}
\end{eqnarray}
and we can isolate $X_{1{+-}}$ through 
\begin{eqnarray}
&& \langle 0 | J_{\mu\nu}^{1{--}} | X_{1{+-}} \rangle \langle X_{1{+-}} | J_{\mu^\prime\nu^\prime}^{1{--}\dagger} | 0 \rangle
\\ \nonumber &=& f_{1{+-}}^2 ~ (q_\mu \epsilon_\nu - q_\nu\epsilon_\mu)(q_{\mu^\prime} \epsilon_{\nu^\prime}^* - q_{\nu^\prime}\epsilon_{\mu^\prime}^*)
\\ \nonumber &=&  f_{1{+-}}^2 ~  ( g_{\mu \mu^\prime}q_\nu q_{\nu^\prime} + g_{\nu \nu^\prime}q_\mu q_{\mu^\prime} 
\\ \nonumber && -g_{\mu \nu^\prime}q_\nu q_{\mu^\prime} - g_{\nu \mu^\prime}q_\mu q_{\nu^\prime} ) \, .
\label{eq:coupling6}
\end{eqnarray}
Similarly, we can investigate the operator \( J_{\mu\nu}^{1{++}} \), which can couple to both \( J^{PC} = 1^{++} \) and \( J^{PC} = 1^{-+} \) meson states.

\item The operator $J_{\mu}^{1{-+}}$ contains both $J^{PC} = 1^{-+}$ and $0^{++}$ components, and therefore it can couple to meson states with these quantum numbers through
\begin{eqnarray}
\langle 0 | J_{\mu}^{1{-+}} | X_{1{-+}} \rangle &=& \epsilon_\mu f_{1{-+}} \, ,
\label{eq:coupling10}
\\ \langle 0 | J_{\mu}^{1{-+}} | X_{0{++}} \rangle &=& q_\mu f_{0{++}} \, ,
\label{eq:coupling11}
\end{eqnarray}
where $f_{1{-+}}$ and $f_{0{++}}$ are the decay constants. These two states, $X_{1{-+}}$ and $X_{0{++}}$, can be clearly separated at the hadron level by analyzing the two-point correlation function
\begin{eqnarray}
&& \Pi^{1{-+}}_{\mu\mu^\prime}(q^2)
\label{eq:coupling12}
\\ \nonumber &\equiv& i \int d^4x e^{iqx} \langle 0 | {\bf T}[J_{\mu}^{1{-+}}(x) J_{\mu^\prime}^{1{-+},\dagger}(0)] | 0 \rangle
\\ \nonumber &=& (g_{\mu\mu^\prime} - q_\mu q_{\mu^\prime}/q^2 )~\Pi_{1{-+}}(q^2) + ( q_\mu q_{\mu^\prime}/q^2 )~ \Pi_{0{++}}(q^2) \, .
\end{eqnarray}
Similarly, we can investigate the operators $J_{\mu}^{1{++}}$, $J_{\mu}^{1{--}}$, and $J_{\mu}^{1{+-}}$.

\end{itemize}

%
%=====================================================================================
%=====================================================================================
\section{QCD sum rule Analysis}
\label{sec:sumrule}
%=====================================================================================
%=====================================================================================
%

The QCD sum rule method represents a well-established and powerful non-perturbative framework in the study of hadron physics, as demonstrated in numerous foundational works~\cite{Shifman:1978bx,Reinders:1984sr,Shifman:1978by,Shifman:1978bw,Novikov:1983gd,Grozin:1994hd,Grozin:2007zz}. In the present work, we employ this formalism to systematically examine the properties of the ten excited light meson operators introduced in Eqs.~(\ref{def:Ja1pm})–(\ref{def:Jm1mp}) and Eq.~(\ref{def:Jm1pp}). For clarity and conciseness, we focus our analysis on the representative case of the \( 2^{++} \) tensor operator \( J_{\mu \nu}^{2{++}} \) defined in Eq.~(\ref{def:Jam2pp}), with particular emphasis on the investigation of its two-point correlation function
%
%%%%%%%%%%%%%%%%%%%%%%%%%%%%%%%%%%%%%%%%%%%%%%%%%%%%%%%%%%%%%%%%%%%%%%%%%%%%%%
\begin{eqnarray}
\nonumber \Pi_{\mu \nu,\mu^\prime \nu^\prime}^{2{++}}(q^2) &=& i \int d^4x e^{iqx} \langle 0 | {\bf T}[ J_{\mu \nu}^{2{++}}(x) J_{\mu^\prime \nu^\prime}^{2{++},\dagger}(0)] | 0 \rangle \\
&=& \Pi_{2{++}}(q^2) \times \epsilon_{\mu\nu} \epsilon^*_{\mu^\prime\nu^\prime} + \cdots \, ,
\label{def:pi}
\end{eqnarray}
%%%%%%%%%%%%%%%%%%%%%%%%%%%%%%%%%%%%%%%%%%%%%%%%%%%%%%%%%%%%%%%%%%%%%%%%%%%%%%
%
at both the hadron and quark-gluon levels, where \( \cdots \) denotes the other structures with different Lorentz structures, such as \( q_\mu q_\nu q_\mu^\prime q_\nu^\prime \), which are omitted as they do not affect the final numerical results. The tensor structure is given by
\begin{equation}
\epsilon_{\mu\nu} \epsilon^*_{\mu^\prime\nu^\prime} = g_{\mu \mu^\prime} g_{\nu \nu^\prime} + g_{\mu \nu^\prime} g_{\nu \mu^\prime} - \frac{1}{2} g_{\mu \nu} g_{\mu^\prime \nu^\prime} \, .
\end{equation}

At the hadron level, the dispersion relation can be derived using Cauchy's contour integration, allowing the correlation function in Eq.~(\ref{def:pi}) to be expressed as:
\begin{equation}
\Pi_{2^{++}}(q^2) = \int_{t_{\min}}^\infty \frac{\rho_{\rm phen}(s)}{s - q^2 - i\varepsilon} \, ds \, ,
\label{eq:hadron}
\end{equation}
where \( t_{\min} = \{M_{2^{++}}^2, s_0^h \} \), with \( M_{2^{++}} \) denoting the mass of the meson \( X_{2^{++}} \), and \( s_0^h \) the threshold of the lowest continuum state. The phenomenological spectral density \( \rho_{\rm phen}(s) \) is parameterized as
\begin{eqnarray}
\rho_{\rm phen}(s) &\equiv& \frac{1}{\pi}\mathrm{Im}\Pi_{2^{++}}(q^2)
\label{eq:rho}
\\ \nonumber
&=& f_{2^{++}}^2 \delta(s - M_{2^{++}}^2) + \rho_{\rm phen}^{\rm higher}(s) \theta(s - s_0^h) \, ,
\end{eqnarray}
where \( \rho_{\rm phen}^{\rm higher}(s) \) accounts for the contributions from higher excited states and the continuum.

At the quark-gluon level, we employ the operator product expansion (OPE) to evaluate the correlation function in Eq.~(\ref{def:pi}), from which the OPE spectral density \( \rho_{\rm OPE}(s) \) is extracted. The correlation function is then expressed as:
\begin{equation}
\Pi_{2^{++}}(q^2) = \int_{s_<}^\infty \frac{\rho_{\rm OPE}(s)}{s - q^2 - i\varepsilon} \, ds \, ,
\label{eq:rhoOPE1}
\end{equation}
where \( s_< \) denotes the physical threshold at the OPE level. Specifically, we take \( s_< = 0 \), \( m_s^2 \), and \( 4m_s^2 \) for the quark contents \( \bar{q}q \), \( \bar{q}s \), and \( \bar{s}s \), respectively (\( q = u, d \)).

Based on the principle of quark-hadron duality—which asserts that in the deep Euclidean limit \( q^2 \to -\infty \), the correlation function becomes dominated by perturbative contributions while condensate effects vanish—we obtain the relation:
\begin{equation}
\int_{s_0^h}^\infty \frac{\rho_{\rm phen}^{\rm higher}(s)}{s(s - q^2)} \, ds = \int_{s_0}^\infty \frac{\rho_{\rm OPE}(s)}{s(s - q^2)} \, ds \, ,
\label{eq:rhoOPEs0}
\end{equation}
where the parameter \( s_0 \) serves as a continuum threshold to be determined through phenomenological analysis. Note that \( s_0 \) and \( s_0^h \) are not necessarily equal.

To enhance the ground-state contribution and suppress that from higher states and the continuum, we perform a Borel transformation on both sides. Combining Eqs.~(\ref{eq:hadron}) and (\ref{eq:rhoOPE1}), we arrive at:
\begin{eqnarray}
&& \int_{s_<}^\infty \rho_{\rm OPE}(s) e^{-s / M_B^2} \, ds 
\label{eq:rhoOPE2}
\\ \nonumber
&=& f_{2^{++}}^2 e^{-M_{2^{++}}^2 / M_B^2} + \int_{s_0^h}^\infty \rho_{\rm phen}^{\rm higher}(s) e^{-s / M_B^2} \, ds \, .
\end{eqnarray}
This yields the QCD sum rule relation:
\begin{eqnarray}
\Pi_{2^{++}}(s_0, M_B^2) &\equiv& f_{2^{++}}^2 \, e^{-M_{2^{++}}^2 / M_B^2}
\label{eq:fin}
\\ \nonumber
&=& \int_{s_<}^{s_0} e^{-s / M_B^2} \rho_{\rm OPE}(s) \, ds \, .
\end{eqnarray}
The mass and decay constant of the meson \( X_{2^{++}} \) are then extracted using:
\begin{eqnarray}
M_{2^{++}}^2(s_0, M_B^2) &=& \frac{ \int_{s_<}^{s_0} e^{-s / M_B^2} s \rho_{\rm OPE}(s) \, ds }{ \int_{s_<}^{s_0} e^{-s / M_B^2} \rho_{\rm OPE}(s) \, ds } \, ,
\label{eq:LSR}
\\
f_{2^{++}}^2(s_0, M_B^2) &=& \Pi_{2^{++}}(s_0, M_B^2) \cdot e^{M_{2^{++}}^2 / M_B^2} \, .
\label{eq:decay}
\end{eqnarray}

\begin{figure*}[]
\begin{center}
\subfigure[(a)]{
\scalebox{0.30}{\includegraphics{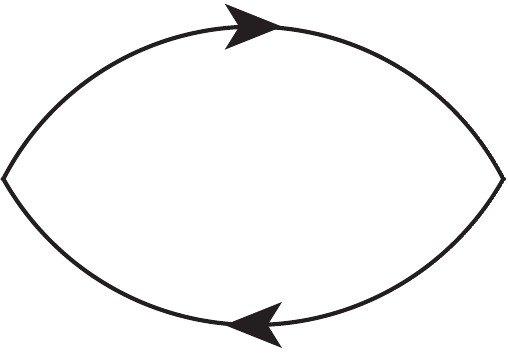}}}
\\[2mm]
\subfigure[(b--1)]{
\scalebox{0.30}{\includegraphics{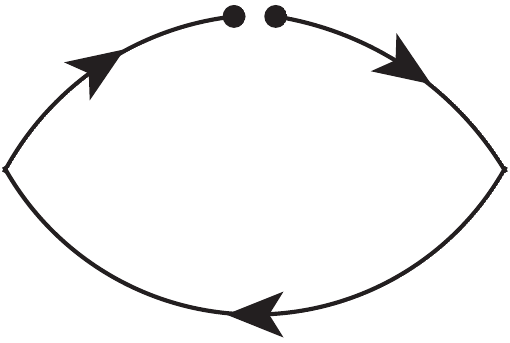}}}~~~~~
\subfigure[(b--2)]{
\scalebox{0.30}{\includegraphics{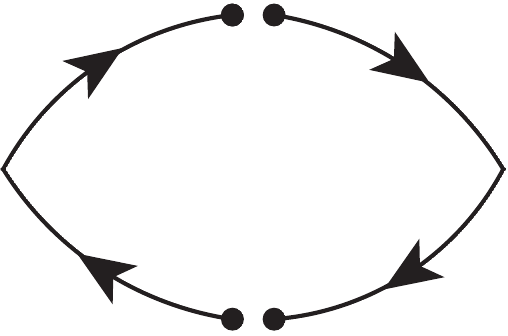}}}~~~~~
\\[2mm]
\subfigure[(c--1)]{
\scalebox{0.30}{\includegraphics{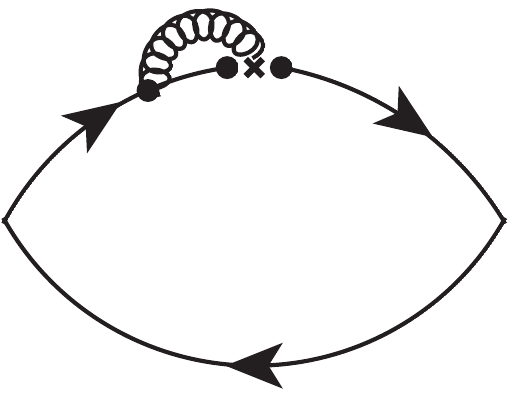}}}~~~~~
\subfigure[(c--2)]{
\scalebox{0.30}{\includegraphics{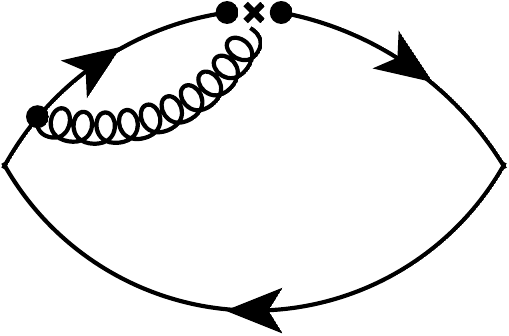}}}~~~~~
\subfigure[(d--1)]{
\scalebox{0.30}{\includegraphics{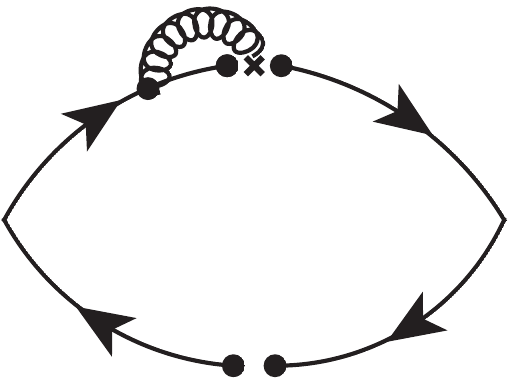}}}~~~~~
\subfigure[(d--2)]{
\scalebox{0.30}{\includegraphics{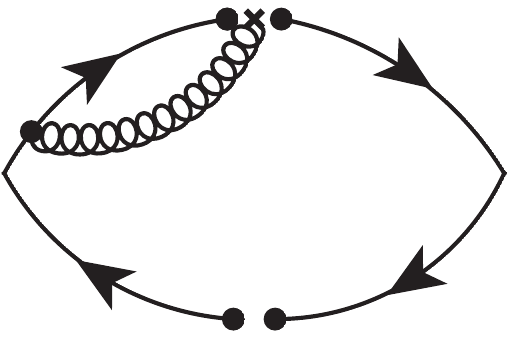}}}~~~~~
\\[2mm]
\subfigure[(e--1)]{
\scalebox{0.30}{\includegraphics{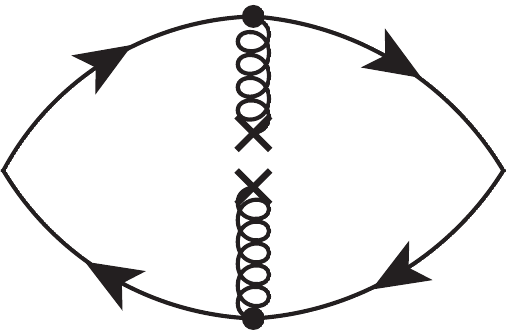}}}~~~~~
\subfigure[(e--2)]{
\scalebox{0.30}{\includegraphics{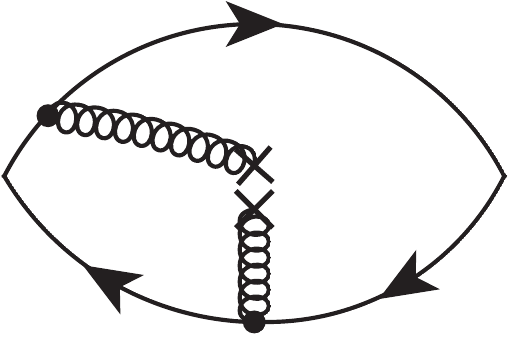}}}~~~~~
\end{center}

\caption{Feynman diagrams for excited light meson operators. Diagrams (a) and (b--i) are proportional to $g_s^{N=0}$; diagrams (c--i) and (d--i) are proportional to $g_s^{N=1}$; and diagrams (e--i) are proportional to $g_s^{N=2}$. The gluons appearing in diagrams (c--2), (d--2), and (e--2) originate (partly) from the covariant derivative operator.}
\label{fig:feynman}
\end{figure*}

Specifically, the sum rule equation derived from the operator $J_{\mu\nu}^{2{++}}$ with the quark content $\bar{q}s$ ($q = u/d$) is given by
\begin{eqnarray}
\Pi_{2{++}} &=& \int^{s_0}_{m_s^2} \Bigg [
- {3 s^2\over 20 \pi^2}
+ \Big( +{m_s^2 \over 2 \pi^2 }+{m_q^2 \over 2 \pi^2 }
\label{eq:piJ1}
\\ \nonumber &&
-{m_q m_s \over 2 \pi^2 }\Big)s
+{\langle g_s^2 GG \rangle \over 18 \pi^2}
-{ m_s^2 m_q^2 \over2\pi^2 } \Bigg ] e^{-s/M_B^2} ds
\\ \nonumber &&
+\Big( +{  m_q \langle g_s \bar s \sigma G s \rangle }
+{  m_s \langle g_s \bar q \sigma G q \rangle } \Big)
\\ \nonumber &&
+ {16 \pi^2 \over M_B^2} \Big(
+{\langle \bar q q \rangle \langle g_s \bar s \sigma G s \rangle \over 48}
+{\langle \bar s s \rangle \langle g_s \bar q \sigma G q \rangle \over 48}
\Big) \, .
\end{eqnarray}
Within the QCD sum rule framework, we derive the sum rule equation for the $\bar{q}q$ quark content through systematic substitutions: $m_s \rightarrow m_q$, $\left\langle {\bar ss} \right\rangle \rightarrow \left\langle {\bar qq} \right\rangle$, and $\left\langle {{g_s}\bar s\sigma Gs} \right\rangle \rightarrow \left\langle {{g_s}\bar q\sigma Gq} \right\rangle$. Conversely, the corresponding expression for the $\bar{s}s$ configuration can be obtained by applying the inverse replacements. Within the QCD sum rule framework, the up and down quarks are treated as degenerate. As a result, for each operator, the extracted results for the isoscalar and isovector $\bar q q$ states are identical. For completeness, we provide the full set of sum rule equations obtained from the remaining operators in Appendix~\ref{app:ope}. 

In the calculations, we take into account the Feynman diagrams depicted in Fig.~\ref{fig:feynman}. The covariant derivative operator \( D_\alpha = \partial_\alpha + i g_s A_\alpha \) can be naturally separated into two terms, with the gluon field from the second term illustrated explicitly in Fig.~\ref{fig:feynman}(c--2), Fig.~\ref{fig:feynman}(d--2), and Fig.~\ref{fig:feynman}(e--2). We consider the perturbative term, the strange quark mass \( m_s \), the quark condensates \( \langle \bar{q} q \rangle \) and \( \langle \bar{s} s \rangle \), the gluon condensate \( \langle g_s^2 GG \rangle \), the quark-gluon mixed condensates \( \langle g_s \bar{q} \sigma G q \rangle \) and \( \langle g_s \bar{s} \sigma G s \rangle \), as well as their combinations. Vacuum saturation is assumed for higher-dimensional condensates. Other condensates, such as \( \langle g_s^3G^3\rangle \) and \( \langle g_s \bar{s} D_\mu G^{\mu\nu} \gamma_\nu s \rangle \), are not considered in the present study. We have accounted for all diagrams proportional to \( g_s^{N=0} \) and \( g_s^{N=1} \), but only partially for those proportional to \( g_s^{N=2} \). Notably, the \( D=6 \) term \( m_s \langle g_s \bar{q} \sigma G q \rangle \), as well as the \( D=8 \) terms \( \langle \bar{q} q \rangle \langle g_s \bar{s} \sigma G s \rangle \) and \( \langle \bar{s} s \rangle \langle g_s \bar{q} \sigma G q \rangle \), play significant roles.

We employ Eq.~(\ref{eq:piJ1}) for the numerical analysis and refer to the corresponding state as \( | {\bar q s}; 2^{+(+)} \rangle \). The following values for various QCD parameters are adopted, as suggested in Refs.~\cite{Colangelo:2000dp,Yang:1993bp,Narison:2002pw,Gimenez:2005nt,Jamin:2002ev,Ioffe:2002be,Ovchinnikov:1988gk,Ellis:1996xc,pdg}:
%
%%%%%%%%%%%%%%%%%%%%%%%%%%%%%%%%%%%%%%%%%%%%%%%%%%%%%%%%%%%%%%%%%%%%%%%%%%%%%%
\begin{eqnarray}
\nonumber m_q &=& m_u = m_d \rightarrow 0 \, ,
\\ \nonumber m_s(2\mbox{ GeV}) &=& 93 ^{+11}_{-5} \mbox{ MeV} \, ,
\\ \nonumber  \langle g_s^2GG\rangle &=& (6.35\pm 0.35) \times 10^{-2} \mbox{ GeV}^4 \, ,
\\ \langle\bar qq \rangle &=& -(0.240 \pm 0.010)^3 \mbox{ GeV}^3 \, ,
\label{condensates}
\\ \nonumber \langle\bar ss \rangle &=& (0.8\pm 0.1)\times \langle\bar qq \rangle \, ,
\\
\nonumber \langle g_s\bar q\sigma G q\rangle &=& - M_0^2\times\langle\bar qq\rangle \, ,
\\
\nonumber \langle g_s\bar s\sigma G s\rangle &=& - M_0^2\times\langle\bar ss\rangle \, ,
\\
\nonumber M_0^2 &=& (0.8 \pm 0.2) \mbox{ GeV}^2 \, .
\end{eqnarray}
%%%%%%%%%%%%%%%%%%%%%%%%%%%%%%%%%%%%%%%%%%%%%%%%%%%%%%%%%%%%%%%%%%%%%%%%%%%%%%
%
According to Eq.~(\ref{eq:LSR}), the mass of the state \( | \bar{q}s; 2^{+(+)} \rangle \) depends on two free parameters: the continuum threshold \( s_0 \) and the Borel mass \( M_B \). To determine their respective working regions, we systematically analyze three key criteria: (a) the convergence of OPE, (b) the dominance of the pole contribution over the continuum, and (c) the stability of the extracted mass with respect to variations in these parameters.

\begin{figure}[hbt]
\begin{center}
\includegraphics[width=0.45\textwidth]{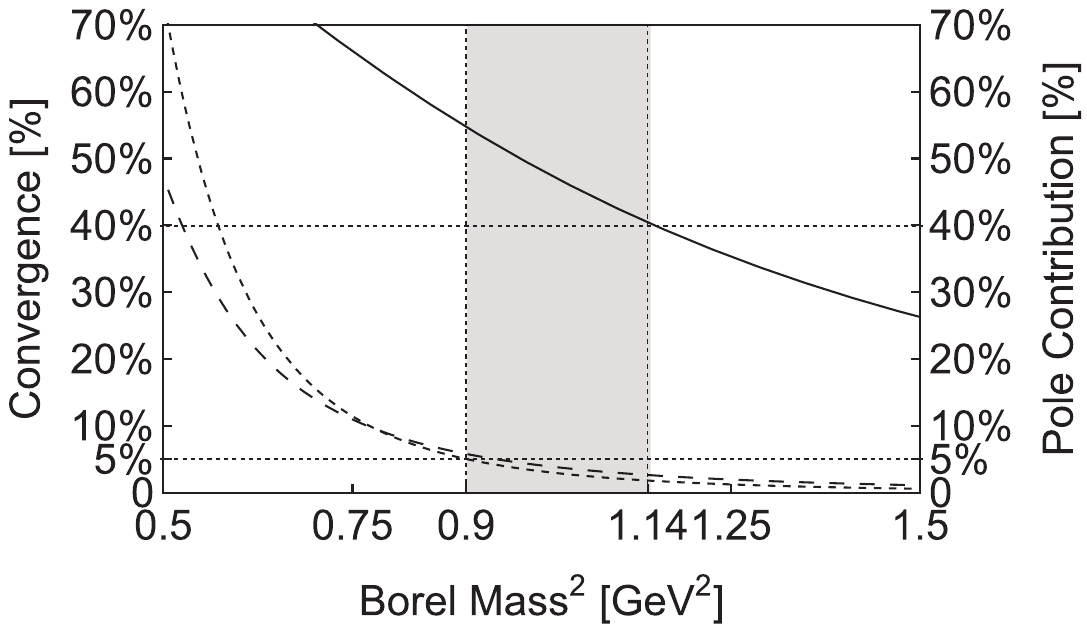}
\caption{CVG$_{8}$ (short dashed curve defined in Eq.~(\ref{eq:convergence8})), CVG$_{6}$ (middle dashed curve defined in Eq.~(\ref{eq:convergence6})), and the pole contribution (solid curve defined in Eq.~(\ref{eq:pole})) as functions of the Borel mass $M_B$. These curves are obtained using the operator $J_{\mu \nu}^{2{++}}$ with the quark content $\bar{q}s$ ($q = u/d$), under the threshold setting $s_0 = 2.9$~GeV$^2$.}
\label{fig:cvgpole}
\end{center}
\end{figure}

\begin{figure*}[]
\begin{center}
\includegraphics[width=0.45\textwidth]{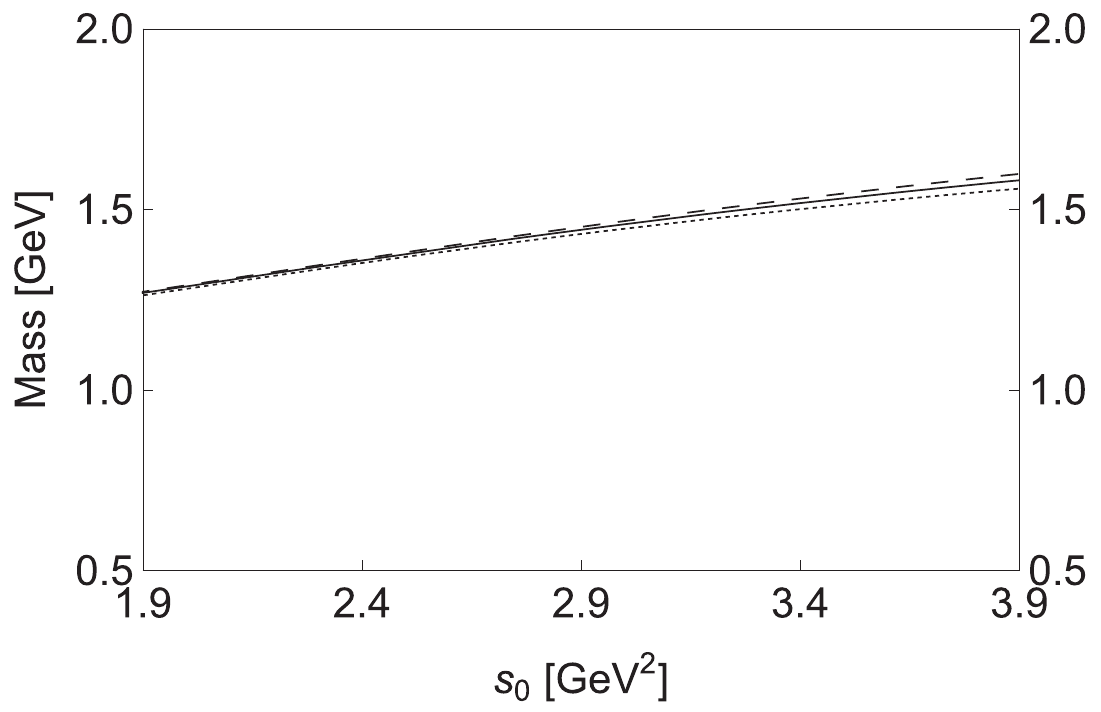}
~~~~~
\includegraphics[width=0.45\textwidth]{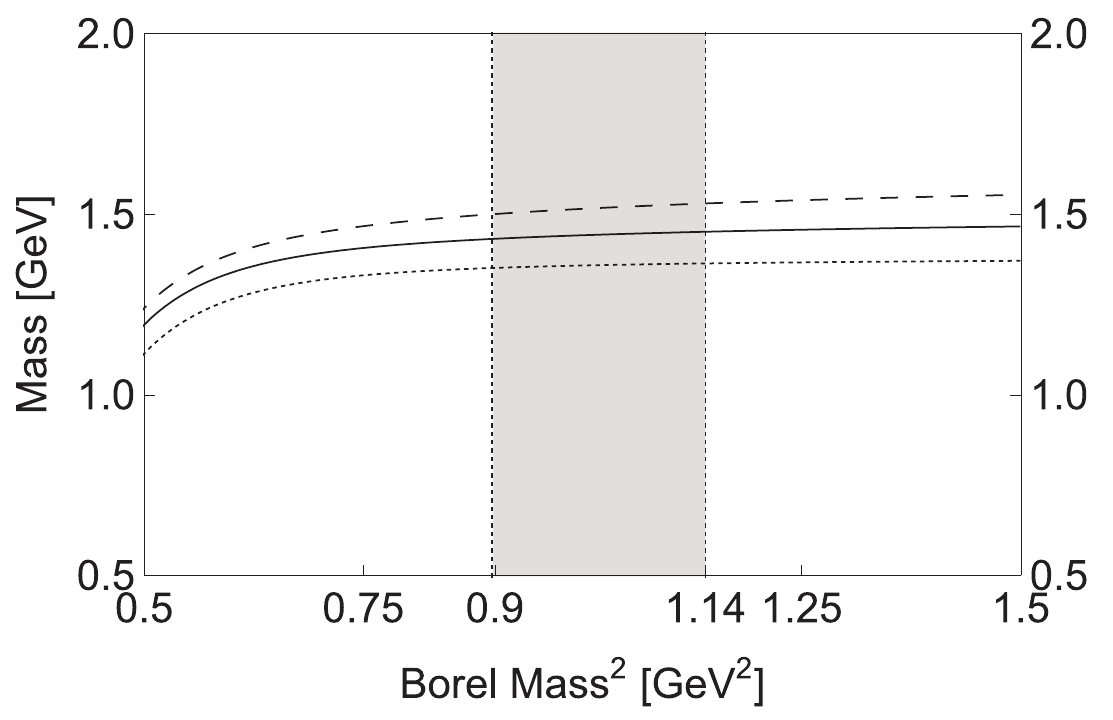}
\caption{
Mass of the state \( | \bar{q}s; 2^{+(+)} \rangle \), extracted from the operator $J_{\mu \nu}^{2{++}}$ with the quark content $\bar{q}s$ ($q = u/d$), as a function of the threshold value $s_0$ (left) and the Borel mass $M_B$ (right).
In the left panel the short-dashed/solid/long-dashed curves are plotted by setting $M_B^2 = 0.90/1.02/1.14$ GeV$^2$, respectively.
In the right panel the short-dashed/solid/long-dashed curves are plotted by setting $s_0 = 2.4/2.9/3.4$ GeV$^2$, respectively.}
\label{fig:J1mass}
\end{center}
\end{figure*}

Firstly, we investigate the convergence of OPE, and require the $D=8$ and $D=6$ terms to be respectively less than 5\% and 10\%:
\begin{eqnarray}
\mbox{CVG}_{8} &\equiv& \left|\frac{ \Pi_{2{++}}^{D=8}(\infty, M_B^2) }{ \Pi_{2{++}}(\infty, M_B^2) }\right| < 5\% \, ,
\label{eq:convergence8}
\\
\mbox{CVG}_{6} &\equiv& \left|\frac{ \Pi_{2{++}}^{D=6}(\infty, M_B^2) }{ \Pi_{2{++}}(\infty, M_B^2) }\right| < 10\% \, .
\label{eq:convergence6}
\end{eqnarray}
As illustrated in Fig.~\ref{fig:cvgpole} (dashed curves), we determine a lower bound for the Borel mass as $M_B^2 > 0.90$~GeV$^2$.

Secondly, we investigate the one-pole-dominance assumption, and require the pole contribution to be larger than 40\%:
\begin{equation}
\mbox{Pole contribution} \equiv \left|\frac{ \Pi_{2{++}}(s_0, M_B^2) }{ \Pi_{2{++}}(\infty, M_B^2) }\right| > 40\% \, .
\label{eq:pole}
\end{equation}
As shown in Fig.~\ref{fig:cvgpole} (solid curve), we determine an upper bound for the Borel mass, \( M_B^2 < 1.14 \)~GeV\(^2\), when adopting a threshold value of \( s_0 = 2.9 \)~GeV\(^2\). Based on this criterion, the valid Borel window at this threshold is identified as \( 0.90 < M_B^2 < 1.14 \)~GeV\(^2\). Extending the analysis to other values of \( s_0 \), we find that viable Borel windows persist as long as the threshold satisfies \( s_0 > s_0^{\text{min}} = 2.4 \)~GeV\(^2\). In practice, \( s_0 \) is typically chosen to be 10\%--20\% larger than \( s_0^{\text{min}} \). In this study, we adopt \( s_0 = 2.9 \)~GeV\(^2\), within the range of \( 2.4 < s_0 < 3.4 \)~GeV\(^2\).

Thirdly, we examine the parameter dependence of the extracted mass, \( M_{| \bar{q}s; 2^{+(+)} \rangle} \), by systematically varying both the threshold value \( s_0 \) and the Borel mass \( M_B \). As shown in Fig.~\ref{fig:J1mass}, our analysis reveals that the mass exhibits moderate sensitivity to \( s_0 \) near 2.9~GeV\(^2\), while demonstrating remarkable stability against variations in \( M_B \) across the Borel window of \( 0.90 \)~GeV\(^2\) \( < M_B^2 < 1.14 \)~GeV\(^2\). Based on these observations, we establish the optimal working regions as \( 2.4 \)~GeV\(^2\) \( < s_0 < 3.4 \)~GeV\(^2\) and \( 0.90 \)~GeV\(^2\) \( < M_B^2 < 1.14 \)~GeV\(^2\). Within these carefully determined parameter ranges, we obtain stable predictions for both the mass and decay constant of the \( | \bar{q}s; 2^{+(+)} \rangle \) state as
\begin{eqnarray}
M_{| \bar{q}s; 2^{+(+)} \rangle} &=& 1.44^{+0.08}_{-0.09} \, \text{GeV} \, ,
\label{eq:qsmass} \\
f_{| \bar{q}s; 2^{+(+)} \rangle} &=& 0.313^{+0.078}_{-0.079} \, \text{GeV}^3 \, .
\end{eqnarray}
Its central value corresponds to $s_0=2.9$~GeV$^2$ and $M_B^2 = 1.02$~GeV$^2$, and its uncertainty comes from the threshold value $s_0$, the Borel mass $M_B$, and various QCD parameters listed in Eqs.~(\ref{condensates}).

We systematically apply the same methodology to analyze the remaining operators given in Eqs.~(\ref{def:Ja1pm})–(\ref{def:Jm1mp}) and Eq.~(\ref{def:Jm1pp}). The comprehensive results of these analyses are summarized in Table~\ref{tab:results}. Notably, for certain cases---the spin-0 states coupled by $J^{0{++}}$ and $J^{0{--}}$, the spin-0 states coupled by $J_{\mu}^{1{--}}$ and $J_{\mu}^{1{+-}}$, as well as the spin-1 states coupled by $J_{\alpha}^{1{-+}}$ and $J_{\alpha}^{1{++}}$---the derived QCD sum rule expressions lack a perturbative term. The absence of this leading-order contribution compromises the reliability of the corresponding predictions. Therefore, we do not report results for these cases.

%
%=====================================================================================
%=====================================================================================
\section{Summary and Discussions}
\label{sec:summary}
%=====================================================================================
%=====================================================================================

\begin{table*}[hpt]
\begin{center}
\renewcommand{\arraystretch}{1.63}
\caption{QCD sum rule results extracted from the excited light meson operators defined in Eqs.~(\ref{def:Ja1pm})–(\ref{def:Jm1mp}) and Eq.~(\ref{def:Jm1pp}), with the quark contents $\bar q q$, $\bar q s$, and $\bar s s$ $(q=u/d)$. For each operator, the extracted results for the isoscalar and isovector $\bar q q$ states are identical. The corresponding experimental candidates, as reported in PDG~\cite{pdg}, are listed in the last column for comparison.}
\resizebox{1.0\textwidth}{!}{ 
\begin{tabular}{c|c|c|c|c|c|c|c|c}
\hline\hline
\multirow{2}{*}{Operators}~&~\multirow{2}{*}{Meson~[$J^{PC}$]}~&~$s_0^{min}$~& \multicolumn{2}{c|}{Working Regions}&~\multirow{2}{*}{Pole~[\%]}~&~\multirow{2}{*}{Mass~[GeV]}~&~\multirow{2}{*}{Decay Constant}~&~\multirow{2}{*}{Candidate~\cite{pdg}}~
\\ \cline{4-5}
&$ $&~~$[{\rm GeV}^2]$~~&~~$M_B^2~[{\rm GeV}^2]$~~&~~$s_0~[{\rm GeV}^2]$~~&&&

\\ \hline\hline
\multirow{4}{*}{$ J_{\mu}^{1{--}}$}  &$| \bar q q; 1^{--}\rangle $          & $3.8$      &   $1.70$--$2.05$   &  $4.6$   &  $40$--$52$  & $1.70^{+0.14}_{-0.17}$  &  $0.602^{+0.173}_{-0.169} $ $\mathrm{GeV^3}$  &$\rho(1700)/\omega(1650)$
\\ \cline{2-9}
& $| \bar q s; 1^{-}\rangle $           & $3.8$              &   $1.71$--$2.09$   &  $4.7$   &  $40$--$53$  & $1.71^{+0.14}_{-0.17}$  &  $0.617^{+0.173}_{-0.170}$ $\mathrm{GeV^3}$  &$K^*(1680)$
\\ \cline{2-9}
& $| \bar s s; 1^{--}\rangle $           & $5.1$              &   $2.25$--$2.81$   &  $6.4$   &  $40$--$55$  & $2.00^{+0.15}_{-0.19}$  &  $0.968^{+0.260}_{-0.257}$ $\mathrm{GeV^3}$  & --
\\ \cline{2-9}
& $| \bar q q/\bar q s/\bar s s; 0^{+(-)}\rangle $ &  \multicolumn{7}{c}{--}
\\ \hline
\multirow{4}{*}{$ J_{\mu}^{1{+-}}$}   & $| \bar q q; 1^{+-}\rangle $  & $2.1$      &   $1.00$--$1.21$   &  $2.6$   &  $40$--$53$  & $1.22^{+0.10}_{-0.13}$  &  $0.257^{+0.063}_{-0.062}$ $\mathrm{GeV^3}$ &$b_1(1235)/h_1(1170)$
\\ \cline{2-9}
& $| \bar q s; 1^{+}\rangle $           & $2.2$              &   $1.01$--$1.23$   &  $2.7$   &  $40$--$53$  & $1.27^{+0.10}_{-0.12}$  &  $0.271^{+0.065}_{-0.064}$ $\mathrm{GeV^3}$  &$K_1(1270)$
\\ \cline{2-9}
& $| \bar s s; 1^{+-}\rangle $           & $2.6$              &   $1.18$--$1.43$   &  $3.2$   &  $40$--$53$  & $1.41^{+0.10}_{-0.12}$  &  $0.348^{+0.086}_{-0.084}$ $\mathrm{GeV^3}$  &$h_1(1415)$
\\ \cline{2-9}
& $| \bar q q/\bar q s/\bar s s; 0^{-(-)}\rangle $ &  \multicolumn{7}{c}{--}
\\ \hline
\multirow{6}{*}{$J_{\mu\nu}^{1{++}}$}   & $| \bar q q; 1^{++}\rangle $           & $2.1$              &   $0.94$--$1.15$   &  $2.6$   &  $40$--$53$  & $1.25^{+0.10}_{-0.13}$  &  $0.111^{+0.010}_{-0.010}$ $\mathrm{GeV^2}$  &$a_1(1260)/f_1(1285)$
\\ \cline{2-9}
& $| \bar q s; 1^{+}\rangle $           & $2.1$              &   $0.92$--$1.13$   &  $2.6$   &  $40$--$53$  & $1.28^{+0.09}_{-0.11}$  &  $0.106^{+0.011}_{-0.012}$ $\mathrm{GeV^2}$  &$K_1(1270)$
\\ \cline{2-9}
& $| \bar s s; 1^{++}\rangle $           & $2.5$              &   $1.08$--$1.33$   &  $3.1$   &  $40$--$53$  & $1.42^{+0.10}_{-0.12}$  &  $0.114^{+0.012}_{-0.014}$ $\mathrm{GeV^2}$  &$f_1(1420)$
\\ \cline{2-9}
& $| \bar q q; 1^{-+}\rangle $        & $3.5$              &   $2.34$--$2.98$   &  $4.3$   &  $40$--$52$  & $1.65^{+0.11}_{-0.13}$  &  $0.328^{+0.062}_{-0.066}$ $\mathrm{GeV^2}$ &$\pi_1(1600)$
\\ \cline{2-9}
& $| \bar q s; 1^{-}\rangle $        & $3.9$              &   $2.65$--$3.35$   &  $4.8$   &  $40$--$52$  & $1.72^{+0.12}_{-0.14}$  &  $0.364^{+0.070}_{-0.075}$ $\mathrm{GeV^2}$ &$K^*(1680)$
\\ \cline{2-9}
& $| \bar s s; 1^{-+}\rangle $        & $4.6$     &   $3.20$--$4.02$   &  $5.7$   &  $40$--$51$  & $1.85^{+0.13}_{-0.15}$  &  $0.431^{+0.077}_{-0.081}$ $\mathrm{GeV^2}$ &$\eta_1(1855)$
\\ \hline
\multirow{3}{*}{$J_{\mu\nu}^{2{++}}$}     & $| \bar q q; 2^{++}\rangle $  & $2.3$    &   $0.93$--$1.14$   &  $2.8$   &  $40$--$53$  & $1.36^{+0.11}_{-0.14}$  &  $0.286^{+0.091}_{-0.092}$ $\mathrm{GeV^3}$ &$a_2(1320)/f_2(1270)$
\\ \cline{2-9}
& $| \bar q s; 2^{+}\rangle $           & $2.4$              &   $0.90$--$1.14$   &  $2.9$   &  $40$--$54$  & $1.44^{+0.08}_{-0.09}$  &  $0.313^{+0.078}_{-0.079}$ $\mathrm{GeV^3}$  &$K_2^*(1430)$
\\ \cline{2-9}
& $| \bar s s; 2^{++}\rangle $           & $2.5$              &   $0.92$--$1.22$   &  $3.1$   &  $40$--$57$  & $1.52^{+0.08}_{-0.08}$  &  $0.358^{+0.096}_{-0.098}$ $\mathrm{GeV^3}$  &$f_2^\prime(1525)$
\\ \hline
\multirow{6}{*}{$J_{\mu\nu}^{1{--}}$}     & $| \bar q q; 1^{--}\rangle $     & $3.7$    &   $1.60$--$1.97$   &  $4.5$   &  $40$--$53$  & $1.70^{+0.12}_{-0.15}$  &  $0.142^{+0.015}_{-0.017}$ $\mathrm{GeV^2}$  &$\rho(1700)/\omega(1650)$
\\ \cline{2-9}
& $| \bar q s; 1^{-}\rangle $           & $3.7$              &   $1.61$--$2.01$   &  $4.6$   &  $40$--$54$  & $1.71^{+0.12}_{-0.15}$  &  $0.143^{+0.014}_{-0.016}$ $\mathrm{GeV^2}$  & --
\\ \cline{2-9}
& $| \bar s s; 1^{--}\rangle $           & $5.0$     &   $2.16$--$2.74$   &  $6.3$   &  $40$--$55$  & $2.00^{+0.15}_{-0.18}$  &  $0.166^{+0.018}_{-0.020}$ $\mathrm{GeV^2}$  & --
\\ \cline{2-9}
& $| \bar q q; 1^{+-}\rangle $       & $3.2$              &   $2.14$--$2.76$   &  $4.0$   &  $40$--$52$  & $1.60^{+0.10}_{-0.12}$  &  $0.305^{+0.057}_{-0.060}$ $\mathrm{GeV^2}$  &$h_1(1595)$
\\ \cline{2-9}
& $| \bar q s; 1^{+}\rangle $        & $3.5$              &   $2.38$--$2.98$   &  $4.3$   &  $40$--$51$  & $1.64^{+0.11}_{-0.13}$  &  $0.327^{+0.063}_{-0.067}$ $\mathrm{GeV^2}$ & --
\\ \cline{2-9}
& $| \bar s s; 1^{+-}\rangle $        & $4.5$     &   $3.10$--$3.94$   &  $5.6$   &  $40$--$52$  & $1.84^{+0.13}_{-0.15}$  &  $0.423^{+0.077}_{-0.081}$ $\mathrm{GeV^2}$ & --
\\ \hline
\multirow{3}{*}{$J_{\mu\nu}^{2{--}}$}      & $| \bar q q; 2^{--}\rangle $         & $3.4$              &   $1.40$--$1.76$   &  $4.2$   &  $40$--$54$  & $1.69^{+0.10}_{-0.12}$  &  $0.567^{+0.147}_{-0.147}$ $\mathrm{GeV^3}$  & --
\\ \cline{2-9}
& $| \bar q s; 2^{-}\rangle $           & $4.1$              &   $1.73$--$2.13$   &  $5.0$   &  $40$--$53$  & $1.81^{+0.13}_{-0.15}$  &  $0.728^{+0.199}_{-0.197}$ $\mathrm{GeV^3}$  &$K_2(1820)$
\\ \cline{2-9}
& $| \bar s s; 2^{--}\rangle $           & $4.5$              &   $1.90$--$2.39$   &  $5.6$   &  $40$--$55$  & $1.91^{+0.13}_{-0.16}$  &  $0.855^{+0.230}_{-0.228}$ $\mathrm{GeV^3}$  & --
\\ \hline
\multirow{4}{*}{$ {J_{\mu}^{1{-+}}}$}   & $| \bar q q; 0^{++}\rangle $      & $2.8$       &   $1.23$--$1.51$   &  $3.4$   &  $40$--$53$  & $1.45^{+0.12}_{-0.14}$  &  $0.310^{+0.030}_{-0.032}$ $\mathrm{GeV^2}$  &$a_0(1450)/f_0(1370)$
\\ \cline{2-9}
& $| \bar q s; 0^{+}\rangle $           & $2.7$              &   $1.20$--$1.46$   &  $3.3$   &  $40$--$53$  & $1.43^{+0.12}_{-0.15}$  &  $0.304^{+0.030}_{-0.032}$ $\mathrm{GeV^2}$  &$K_0^*(1430)$
\\ \cline{2-9}
& $| \bar s s; 0^{++}\rangle $           & $4.0$              &   $1.72$--$2.12$   &  $4.9$   &  $40$--$53$  & $1.78^{+0.13}_{-0.16}$  &  $0.355^{+0.039}_{-0.044}$ $\mathrm{GeV^2}$  &$f_0(1770)$
\\ \cline{2-9}
& $| \bar q q/\bar q s/\bar s s; 1^{-(+)}\rangle $ &  \multicolumn{7}{c}{--}
\\ \hline
\multirow{4}{*}{$ {J_{\mu}^{1{++}}}$ }  & $| \bar q q; 0^{-+}\rangle $   & $4.1$           &   $1.80$--$2.19$   &  $5.0$   &  $40$--$53$  & $1.78^{+0.13}_{-0.16}$  &  $0.368^{+0.037}_{-0.042}$ $\mathrm{GeV^2}$ &$\pi(1800)/\eta(1760)$ 
\\ \cline{2-9}
& $| \bar q s; 0^{-}\rangle $           & $4.2$              &   $1.84$--$2.28$   &  $5.2$   &  $40$--$54$  & $1.83^{+0.13}_{-0.15}$  &  $0.372^{+0.037}_{-0.042}$ $\mathrm{GeV^2}$  &$K(1830)$
\\ \cline{2-9}
& $| \bar s s; 0^{-+}\rangle $           & $5.1$              &   $2.18$--$2.76$   &  $6.4$   &  $40$--$55$  & $2.03^{+0.15}_{-0.18}$  &  $0.405^{+0.044}_{-0.050}$ $\mathrm{GeV^2}$  & --
\\ \cline{2-9}
& $| \bar q q/\bar q s/\bar s s; 1^{+(+)}\rangle $ &  \multicolumn{7}{c}{--}
\\ \hline\hline

\end{tabular}
}
\label{tab:results}
\end{center}
\end{table*} 

In this study we employ the QCD sum rule method to systematically investigate excited light meson operators composed of one quark field, one antiquark field, and one covariant derivative. A comprehensive set of twelve such operators is constructed, of which ten are selected for detailed QCD sum rule analyses. We compute the masses and decay constants of the corresponding meson states, considering quark contents of the form \( \bar{q}q \), \( \bar{s}s \), and \( \bar{q}s \) (\( q = u/d \)). The numerical results are summarized in Table~\ref{tab:results}, and are in excellent agreement with the lattice QCD calculations reported in Ref.~\cite{Dudek:2013yja}, thereby substantially reinforcing the credibility of our conclusions. A detailed discussion of each case is presented below (\( \mu/\nu = 0 \cdots 3 \) and \( i/j = 1 \cdots 3 \)):
\begin{itemize}

\item The quark-antiquark pair $\bar q_a \gamma_5 q_a$ carries $L=0$, $S=0$, and $J=0$. By incorporating a covariant derivative, the operator
\begin{equation}
J_{\mu}^{1{+-}} = \bar q_a {\overset{\leftrightarrow}{D}}_\mu\gamma_5 q_a \, ,
\end{equation}
acquires $L=1$, $S=0$, and $J=1$, thereby allowing the investigation of excited $^1P_1$ meson states with quantum numbers $J^{PC}=1^{+-}$ through the coupling
\begin{equation}
\langle 0 | J_{\mu}^{1{+-}} | X_{1{+-}} \rangle = \epsilon_\mu f_{1{+-}} \, .
\end{equation}
Our analysis indicates that the $b_1(1235)$, $h_1(1170)$, $K_1(1270)$, and $h_1(1415)$ can be classified as members of an $SU(3)$ flavor nonet with quantum numbers $J^{P(C)}=1^{+(-)}$. 

\item The quark-antiquark pair $\bar q_a \sigma_{0i} q_a$ carries $L=0$, $S=1$, and $J=1$. By incorporating a covariant derivative, the $\mu \rightarrow 0$ component of the operator
\begin{equation}
J_{\alpha}^{1{-+}} = \bar q_a {\overset{\leftrightarrow}{D}}_\mu \sigma_{\alpha\beta}  q_a \times g_{\mu\beta} \, ,
\end{equation}
acquires $L=1$, $S=1$, and $J=0$, thereby allowing the investigation of excited $^3P_0$ meson states with quantum numbers $J^{PC}=0^{++}$ through the coupling
\begin{equation}
\langle 0 | J_{\alpha}^{1{-+}} | X_{0{++}} \rangle = q_\mu f_{0{++}} \, .
\end{equation}
Our analysis indicates that the $a_0(1450)$, $f_0(1370)$, $K_0^*(1430)$, and $f_0(1770)$ can be classified as members of an $SU(3)$ flavor nonet with quantum numbers $J^{P(C)}=0^{+(+)}$.

\item The quark-antiquark pair $\bar q_a \gamma_i q_a$ carries $L=0$, $S=1$, and $J=1$. By incorporating a covariant derivative, the $\mu\nu \rightarrow ij$ component of the operator
\begin{equation}
J_{\mu\nu}^{1{++}} =  \mathcal{A}[\bar q_a {\overset{\leftrightarrow}{D}}_\mu\gamma_{\nu}  q_a ]\, .
\end{equation}
acquires $L=1$, $S=1$, and $J=1$, thereby allowing the investigation of excited $^3P_1$ meson states with quantum numbers $J^{PC}=1^{++}$ through the coupling
\begin{equation}
\langle 0 | J_{\mu\nu}^{1{++}} | X_{1{++}} \rangle = i f_{1{++}} \epsilon_{\mu\nu \alpha \beta} \epsilon^\alpha q^\beta \, .
\end{equation}
Our analysis indicates that the $a_1(1260)$, $f_1(1285)$, $K_1(1270)$, and $f_1(1420)$ can be classified as members of an $SU(3)$ flavor nonet with quantum numbers $J^{P(C)}=1^{+(+)}$. This conclusion is well consistent with the lattice QCD calculations reported in Ref.~\cite{Dudek:2013yja}.

In addition, the $\mu\nu \rightarrow 0i$ component of the operator $J_{\mu\nu}^{1{++}}$ can be employed to investigate meson states with the exotic quantum numbers $J^{PC}=1^{-+}$ via the coupling
\begin{equation}
\langle 0 | J_{\mu\nu}^{1{++}} | X_{1{-+}} \rangle = i f_{1{-+}} (q_\mu \epsilon_\nu - q_\nu\epsilon_\mu) \, .
\end{equation}
Given that the covariant derivative includes a gluon field, our analysis suggests that the $\pi_1(1600)$ and $\eta_1(1855)$ can be interpreted as hybrid mesons with quantum numbers $J^{PC} = 1^{-+}$.

\item For the quark-antiquark pair $\bar q_a \gamma_i q_a$, the inclusion of a covariant derivative also allows the construction of the operator
\begin{equation}
J_{\mu\nu}^{2{++}} =  \mathcal{S}[\bar q_a {\overset{\leftrightarrow}{D}}_\mu\gamma_{\nu}  q_a ]\, .
\end{equation}
which carries $L=1$, $S=1$, and $J=2$, thereby allowing the investigation of excited $^3P_2$ meson states with quantum numbers $J^{PC}=2^{++}$ through the coupling
\begin{equation}
\langle 0 | J_{\mu\nu}^{2{++}} | X_{2{++}} \rangle = i f_{2{++}} \epsilon_{\mu\nu } \, .
\end{equation}
Our analysis indicates that the \( a_2(1320) \), \( f_2(1270) \), \( K_2^*(1430) \), and \( f_2^\prime(1525) \) can be classified as members of an \( SU(3) \) flavor nonet with quantum numbers \( J^{P(C)} = 2^{+(+)} \).  This conclusion is also well consistent with the lattice QCD calculations reported in Ref.~\cite{Dudek:2013yja}.

\item The quark-antiquark pair $\bar q_a q_a$ carries $L=1$, $S=1$, and $J=0$. By incorporating a covariant derivative, the $\mu \rightarrow i$ component of the operator 
\begin{equation}
J_{\mu}^{1{--}} = \bar q_a {\overset{\leftrightarrow}{D}}_\mu q_a \, ,
\end{equation}
acquires $L=2$, $S=1$, and $J=1$. Similarly, the quark-antiquark pair $\bar q_a \gamma_i\gamma_5 q_a$ carries $L=1$, $S=1$, and $J=1$. By incorporating a covariant derivative, the $\mu\nu \rightarrow ij$ component of the operators 
\begin{equation}
J_{\mu\nu}^{1{--}} = \mathcal{A}[\bar q_a {\overset{\leftrightarrow}{D}}_\mu\gamma_{\nu}\gamma_5  q_a ]\, .
\end{equation}
also acquires $L=2$, $S=1$, and $J=1$. These two operators can thus be used to investigate excited $^3D_1$ meson states with quantum numbers $J^{PC}=1^{--}$, through the couplings
\begin{eqnarray}
\langle 0 | J_{\mu}^{1{--}} | X_{1{--}} \rangle &=& \epsilon_\mu f_{1{--}} \, ,
\\
\langle 0 | J_{\mu\nu}^{1{--}} | X_{1{--}} \rangle &=& i f_{1{--}} \epsilon_{\mu\nu \alpha \beta} \epsilon^\alpha q^\beta \, .
\end{eqnarray}
Our analysis indicates that the $\rho(1700)$, $\omega(1650)$, and $K^*(1680)$ can be classified as members of an $SU(3)$ flavor nonet with quantum numbers $J^{P(C)}=1^{-(-)}$. The corresponding $\bar s s$ state has not yet been observed experimentally, and its mass is calculated to be $2.00^{+0.15}_{-0.19}$~GeV.

In addition, the quark-antiquark pair $\bar q_a \gamma_0\gamma_5 q_a$ carries $L=0$, $S=0$, and $J=0$. By incorporating a covariant derivative, the $\mu\nu \rightarrow 0i$ component of the operator $J_{\mu\nu}^{1{--}}$ acquires $L=1$, $S=0$, and $J=1$. Our results suggest that this operator can be employed to investigate the excited $2^1P_1$ meson state with quantum numbers $J^{PC}=1^{+-}$ through the coupling
\begin{equation}
\langle 0 | J_{\mu\nu}^{1{--}} | X_{1{+-}} \rangle = i f_{1{+-}} (q_\mu \epsilon_\nu - q_\nu\epsilon_\mu) \, .
\end{equation}
Accordingly, the $h_1(1595)$ can be assigned as a member of an $SU(3)$ flavor nonet with quantum numbers $J^{P}=1^{+-}$. The corresponding states with quark contents $\bar q s$ and $\bar s s$ have not yet been observed experimentally, and their masses are calculated to be $1.64^{+0.11}_{-0.13}$~GeV and $1.84^{+0.13}_{-0.15}$~GeV, respectively.

\item For the quark-antiquark pair $\bar q_a \gamma_i\gamma_5 q_a$, the inclusion of a covariant derivative also allows the construction of the operator
\begin{equation}
J_{\mu\nu}^{2{--}} =  \mathcal{S}[\bar q_a {\overset{\leftrightarrow}{D}}_\mu\gamma_{\nu}\gamma_5  q_a ]\, .
\end{equation}
which carries $L=2$, $S=1$, and $J=2$, thereby allowing the investigation of excited $^3D_2$ meson states with quantum numbers $J^{PC}=2{--}$ through the coupling
\begin{equation}
\langle 0 | J_{\mu\nu}^{2{--}} | X_{2{--}} \rangle = i f_{2{--}} \epsilon_{\mu\nu } \, .
\end{equation}
Our analysis indicates that the $K_2(1820)$ can be classified as a member of an $SU(3)$ flavor nonet with quantum numbers $J^{P}=2^{-}$. The corresponding states with quark contents $\bar q q$ and $\bar s s$ have not yet been observed experimentally, and their masses are calculated to be $1.69^{+0.10}_{-0.12}$~GeV and $1.91^{+0.13}_{-0.16}$~GeV, respectively.

\item The quark-antiquark pair $\bar q_a \sigma_{0i}\gamma_5 q_a$ carries $L=1$, $S=0$, and $J=1$. By incorporating a covariant derivative, the $\mu \rightarrow 0$ component of the operator
\begin{equation}
J_{\alpha}^{1{-+}} = \bar q_a {\overset{\leftrightarrow}{D}}_\mu \sigma_{\alpha\beta}\gamma_5  q_a \times g_{\mu\beta}\, ,
\end{equation}
acquires $L=0$, $S=0$, and $J=0$, thereby allowing the investigation of pseudoscalar meson states with quantum numbers $J^{PC} = 0^{-+}$ through the coupling
\begin{equation}
\langle 0 | J_{\alpha}^{1{++}} | X_{0{-+}} \rangle = q_\mu f_{0{-+}} \, .
\end{equation}
Our analysis reveals an unconventional behavior: this operator does not significantly couple to the ground state or the first radial excitation ({\it i.e.}, the $1^1S_0$ or $2^1S_0$ states), but predominantly couples to the $3^1S_0$ state. Accordingly, the $\pi(1800)$, $\eta(1760)$, and $K(1830)$ can be interpreted as members of an $SU(3)$ flavor nonet with quantum numbers $J^{P(C)} = 0^{-(+)}$. The corresponding $\bar s s$ state has not yet been observed experimentally, and its mass is calculated to be $2.03^{+0.15}_{-0.18}$~GeV.

\end{itemize}

%=====================================================================================
%=====================================================================================
%=====================================================================================
\section*{Acknowledgments}
%=====================================================================================
%=====================================================================================
%=====================================================================================
%

We thank Niu Su for useful discussions.
This project is supported by
the National Natural Science Foundation of China under Grant No.~12075019,
the Jiangsu Provincial Double-Innovation Program under Grant No.~JSSCRC2021488,
the SEU Innovation Capability Enhancement Plan for Doctoral Students No.~CXJHSEU25139,
and
the Fundamental Research Funds for the Central Universities.

\appendix

\begin{widetext}
\section{Spectral densities}
\label{app:ope}

In this appendix we present the QCD sum rule equations derived from the operators $J^{\cdots}_{\mu(\nu)}$ defined in Eqs.~(\ref{def:Ja1mm}-\ref{def:Jm1mp}) and Eq.~(\ref{def:Jm1pp}), as follows:
\begin{itemize}

\item The operator $J_{\mu}^{1{--}}$ couples to both $J^{PC} = 1^{--}$ and $0^{+-}$ states. The corresponding QCD sum rule equations for the quark content $\bar{q}s$ ($q = u/d$) are
\begin{eqnarray}
%%%%%%%%%%%%%%%%%%%%%%%%%%%%%%%%%%%%%%%%%%%%%%%%%%%%%%%%%%%%%%%%%%%%%%%%%%%%%%%
%------------------------------\J1----------------------------------
\nonumber
\Pi_{1{--}} &=& \int^{s_0}_{m_s^2} \Bigg [
+ { s^2\over 8 \pi^2}
+ \Big( -{m_s^2 \over 2 \pi^2 }-{m_q^2 \over 2 \pi^2 }-{m_q m_s \over 4 \pi^2 }\Big)s+{\langle g_s^2 GG \rangle \over 32 \pi^2}
+{3 m_s^2 m_q^2 \over4\pi^2 } \Bigg ] e^{-s/M^2} ds
\\  &&
+\Big( +{  m_q \langle g_s \bar s \sigma G s \rangle \over 2}
+{  m_s \langle g_s \bar q \sigma G q \rangle \over 2} \Big)
+ {16 \pi^2 \over M_B^2} \Big(+{ \langle \bar q q \rangle m_q \langle \bar s s \rangle m_s \over 96}
+{\langle \bar q q \rangle \langle g_s \bar s \sigma G s \rangle \over 96}
+{\langle \bar s s \rangle \langle g_s \bar q \sigma G q \rangle \over 96}
 \Big) \, ,
%%%%%%%%%%%%%%%%%%%%%%%%%%%%%%%%%%%%%%%%%%%%%%%%%%%%%%%%%%%%%%%%%%%%%%%%%%%%%%%
%------------------------------\J2---------------------------------
\\\nonumber
{{\Pi_{0{+-}}}} &=& \int^{s_0}_{m_s^2} \Bigg [
- {3 m_q \langle \bar q q \rangle \over2 } - {3 m_s \langle \bar s s \rangle \over2 }- { m_q \langle \bar s s \rangle }
- { m_s \langle \bar q q \rangle }  
 \Bigg ] e^{-s/M^2} ds
+\Big(
+{ m_s m_q^2 \langle \bar s s \rangle  }
+{ m_q m_s^2 \langle \bar q q \rangle  }
\\  &&
-{ m_q \langle g_s \bar s \sigma G s \rangle \over 2 }
-{ m_s \langle g_s \bar q \sigma G q \rangle \over 2 }
\Big) 
+ {16 \pi^2 \over M_B^2} \Big(  
+{  \langle \bar q q \rangle m_q \langle \bar s s \rangle m_s \over 96}
+{\langle \bar q q \rangle \langle g_s \bar s \sigma G s \rangle \over 96}
+{\langle \bar s s \rangle \langle g_s \bar q \sigma G q \rangle \over 96}
\Big) \, .
%%%%%%%%%%%%%%%%%%%%%%%%%%%%%%%%%%%%%%%%%%%%%%%%%%%%%%%%%%%%%%%%%%%%%%%%%%%%%%%
%------------------------------\J3----------------------------------
\end{eqnarray}

\item The operator $J_{\mu}^{1{+-}}$ couples to both $J^{PC} = 1^{+-}$ and $0^{--}$ states. The corresponding QCD sum rule equations for the quark content $\bar{q}s$ ($q = u/d$) are
\begin{eqnarray}
\nonumber
\Pi_{1{+-}} &=& \int^{s_0}_{m_s^2} \Bigg [
+ { s^2\over 8 \pi^2}
+ \Big( -{m_s^2 \over 2 \pi^2 }-{m_q^2 \over 2 \pi^2 }+{m_q m_s \over 4 \pi^2 }\Big)s+{\langle g_s^2 GG \rangle \over 32 \pi^2}
+{3 m_s^2 m_q^2 \over4\pi^2 } \Bigg ] e^{-s/M^2} ds
\\  &&
+\Big( -{  m_q \langle g_s \bar s \sigma G s \rangle \over 2}
-{  m_s \langle g_s \bar q \sigma G q \rangle \over 2} \Big)
+ {16 \pi^2 \over M_B^2} \Big(+{ \langle \bar q q \rangle m_q \langle \bar s s \rangle m_s \over 96}
-{\langle \bar q q \rangle \langle g_s \bar s \sigma G s \rangle \over 96}
-{\langle \bar s s \rangle \langle g_s \bar q \sigma G q \rangle \over 96}
\Big) \, ,
%%%%%%%%%%%%%%%%%%%%%%%%%%%%%%%%%%%%%%%%%%%%%%%%%%%%%%%%%%%%%%%%%%%%%%%%%%%%%%%
%------------------------------\J4---------------------------------
\\\nonumber
{\Pi_{0{--}}} &=& \int^{s_0}_{m_s^2} \Bigg [
- {3 m_q \langle \bar q q \rangle \over2 } - {3 m_s \langle \bar s s \rangle \over2 }+ { m_q \langle \bar s s \rangle }
+ { m_s \langle \bar q q \rangle }  
 \Bigg ] e^{-s/M^2} ds
+\Big(
+{ m_s m_q^2 \langle \bar s s \rangle  }
+{ m_q m_s^2 \langle \bar q q \rangle  }
\\  &&
+{ m_q \langle g_s \bar s \sigma G s \rangle \over 2 }
+{ m_s \langle g_s \bar q \sigma G q \rangle \over 2 }
\Big) 
+ {16 \pi^2 \over M_B^2} \Big(  
+{  \langle \bar q q \rangle m_q \langle \bar s s \rangle m_s \over 96}
-{\langle \bar q q \rangle \langle g_s \bar s \sigma G s \rangle \over 96}
-{\langle \bar s s \rangle \langle g_s \bar q \sigma G q \rangle \over 96}
\Big) \, .
%%%%%%%%%%%%%%%%%%%%%%%%%%%%%%%%%%%%%%%%%%%%%%%%%%%%%%%%%%%%%%%%%%%%%%%%%%%%%%%
%------------------------------\J5----------------------------------
\end{eqnarray}

\item The operator $J_{\mu}^{1{++}}$ couples to both $J^{PC} = 1^{++}$ and $1^{-+}$ states. The corresponding QCD sum rule equations for the quark content $\bar{q}s$ ($q = u/d$) are
\begin{eqnarray}
\nonumber
\Pi_{1{++}} &=& \int^{s_0}_{m_s^2} \Bigg [
+ { s^2\over 16 \pi^2}
+ \Big( -{m_s^2 \over 4 \pi^2 }-{m_q^2 \over 4 \pi^2 }+{m_q m_s \over 8 \pi^2 }\Big)s 
+{3 m_s^2 m_q^2 \over8\pi^2 } \Bigg ] e^{-s/M^2} ds
+\Big( -{  m_q \langle g_s \bar s \sigma G s \rangle \over 4}
\\  &&
-{  m_s \langle g_s \bar q \sigma G q \rangle \over 4} \Big)
+ {16 \pi^2 \over M_B^2} \Big(+{ \langle \bar q q \rangle m_q \langle \bar s s \rangle m_s \over 96}
-{\langle \bar q q \rangle \langle g_s \bar s \sigma G s \rangle \over 192}
-{\langle \bar s s \rangle \langle g_s \bar q \sigma G q \rangle \over 192}
\Big) \, ,
%%%%%%%%%%%%%%%%%%%%%%%%%%%%%%%%%%%%%%%%%%%%%%%%%%%%%%%%%%%%%%%%%%%%%%%%%%%%%%%
%------------------------------\J6----------------------------------
\\ \nonumber
{\Pi_{1{-+}}} &=& \int^{s_0}_{m_s^2} \Big(
+ { s \over 8 \pi^2}
- {15 m_s^2 \over 32 \pi^2 } - {15 m_q^2 \over 32 \pi^2 } +{3 m_q m_s \over 16 \pi^2 }\Big)  e^{-s/M^2} ds
+\Big( - {3 m_s^2 m_q^2 \over4\pi^2 } - { m_q \langle \bar q q \rangle \over2 } - { m_s \langle \bar s s \rangle \over2 }
\\ \nonumber &&
+ { m_q \langle \bar s s \rangle \over4 }
+ { m_s \langle \bar q q \rangle \over4 } - {\langle g_s^2 GG \rangle \over 192} \Big)
+ {1 \over M_B^2} \Big(
+{3 \langle \bar s s \rangle m_q^2  m_s \over 8}
+{3 \langle \bar q q \rangle m_s^2  m_q \over 8}
+{m_q \langle g_s \bar s \sigma G s \rangle \over 2}
\\  &&
+{m_s \langle g_s \bar q \sigma G q \rangle \over 2}
\Big) - {8 \pi^2 \over M_B^4} \Big(  
+{  \langle \bar q q \rangle m_q \langle \bar s s \rangle m_s \over 96}
-{\langle \bar q q \rangle \langle g_s \bar s \sigma G s \rangle \over 192}
-{\langle \bar s s \rangle \langle g_s \bar q \sigma G q \rangle \over 192}
\Big) \, .
%%%%%%%%%%%%%%%%%%%%%%%%%%%%%%%%%%%%%%%%%%%%%%%%%%%%%%%%%%%%%%%%%%%%%%%%%%%%%%%
%------------------------------\J7----------------------------------
\end{eqnarray}

\item The operator $J_{\mu\nu}^{2{++}}$ only couples to the $J^{PC} = 2^{++}$ state. The corresponding QCD sum rule equation for the quark content $\bar{q}s$ ($q = u/d$) is
\begin{eqnarray}
\nonumber
\Pi_{2{++}} &=& \int^{s_0}_{m_s^2} \Bigg [
- {3 s^2\over 20 \pi^2}
+ \Big( +{m_s^2 \over 2 \pi^2 }+{m_q^2 \over 2 \pi^2 }-{m_q m_s \over 2 \pi^2 }\Big)s+{\langle g_s^2 GG \rangle \over 18 \pi^2}
-{ m_s^2 m_q^2 \over2\pi^2 } \Bigg ] e^{-s/M^2} ds
\\  &&
+\Big( +{  m_q \langle g_s \bar s \sigma G s \rangle }
+{  m_s \langle g_s \bar q \sigma G q \rangle } \Big)
+ {16 \pi^2 \over M_B^2} \Big(
+{\langle \bar q q \rangle \langle g_s \bar s \sigma G s \rangle \over 48}
+{\langle \bar s s \rangle \langle g_s \bar q \sigma G q \rangle \over 48}
\Big) \, .
%%%%%%%%%%%%%%%%%%%%%%%%%%%%%%%%%%%%%%%%%%%%%%%%%%%%%%%%%%%%%%%%%%%%%%%%%%%%%%%
%------------------------------\J8----------------------------------
\end{eqnarray}

\item The operator $J_{\mu\nu}^{1{--}}$ couples to both $J^{PC} = 1^{--}$ and $1^{+-}$ states. The corresponding QCD sum rule equations for the quark content $\bar{q}s$ ($q = u/d$) are
\begin{eqnarray}
\nonumber
\Pi_{1{--}} &=& \int^{s_0}_{m_s^2} \Bigg [
+ { s^2\over 16 \pi^2}
+ \Big( -{m_s^2 \over 4 \pi^2 } -{m_q^2 \over 4 \pi^2 } -{m_q m_s \over 8 \pi^2 }\Big)s 
+{3 m_s^2 m_q^2 \over8\pi^2 } \Bigg ] e^{-s/M^2} ds
+\Big( +{  m_q \langle g_s \bar s \sigma G s \rangle \over 4}
\\  &&
+{  m_s \langle g_s \bar q \sigma G q \rangle \over 4} \Big)
+ {16 \pi^2 \over M_B^2} \Big(
+{ \langle \bar q q \rangle m_q \langle \bar s s \rangle m_s \over 96}
+{\langle \bar q q \rangle \langle g_s \bar s \sigma G s \rangle \over 192}
+{\langle \bar s s \rangle \langle g_s \bar q \sigma G q \rangle \over 192}
\Big) \, ,
%%%%%%%%%%%%%%%%%%%%%%%%%%%%%%%%%%%%%%%%%%%%%%%%%%%%%%%%%%%%%%%%%%%%%%%%%%%%%%%
%------------------------------\J9----------------------------------
\\\nonumber
{\Pi_{1{+-}}} &=& \int^{s_0}_{m_s^2} \Big(
+ { s \over 8 \pi^2}
- {15 m_s^2 \over 32 \pi^2 } - {15 m_q^2 \over 32 \pi^2 } -{3 m_q m_s \over 16 \pi^2 }\Big)  e^{-s/M^2} ds
+\Big( - {3 m_s^2 m_q^2 \over4\pi^2 } - { m_q \langle \bar q q \rangle \over2 } - { m_s \langle \bar s s \rangle \over2 }
\\ \nonumber &&
- { m_q \langle \bar s s \rangle \over4 }
- { m_s \langle \bar q q \rangle \over4 } - {\langle g_s^2 GG \rangle \over 192} \Big)
+ {1 \over M_B^2} \Big(
+{3 \langle \bar s s \rangle m_q^2  m_s \over 8}
+{3 \langle \bar q q \rangle m_s^2  m_q \over 8}
-{m_q \langle g_s \bar s \sigma G s \rangle \over 2}
\\  &&
-{m_s \langle g_s \bar q \sigma G q \rangle \over 2}
\Big) - {8 \pi^2 \over M_B^4} \Big(  
+{  \langle \bar q q \rangle m_q \langle \bar s s \rangle m_s \over 96}
+{\langle \bar q q \rangle \langle g_s \bar s \sigma G s \rangle \over 192}
+{\langle \bar s s \rangle \langle g_s \bar q \sigma G q \rangle \over 192}
\Big) \, .
%%%%%%%%%%%%%%%%%%%%%%%%%%%%%%%%%%%%%%%%%%%%%%%%%%%%%%%%%%%%%%%%%%%%%%%%%%%%%%%
%------------------------------\J10---------------------------------
\end{eqnarray}

\item The operator $J_{\mu\nu}^{2{--}}$ only couples to the $J^{PC} = 2^{--}$ state. The corresponding QCD sum rule equation for the quark content $\bar{q}s$ ($q = u/d$) is
\begin{eqnarray}
\nonumber
\Pi_{2{--}} &=& \int^{s_0}_{m_s^2} \Bigg [
- {3 s^2\over 20 \pi^2}
+ \Big( +{m_s^2 \over 2 \pi^2 }+{m_q^2 \over 2 \pi^2 }+{m_q m_s \over 2 \pi^2 }\Big)s+{\langle g_s^2 GG \rangle \over 18 \pi^2}
-{ m_s^2 m_q^2 \over2\pi^2 } \Bigg ] e^{-s/M^2} ds
\\  &&
+\Big( -{  m_q \langle g_s \bar s \sigma G s \rangle }
-{  m_s \langle g_s \bar q \sigma G q \rangle } \Big)
+ {16 \pi^2 \over M_B^2} \Big(
-{\langle \bar q q \rangle \langle g_s \bar s \sigma G s \rangle \over 48}
-{\langle \bar s s \rangle \langle g_s \bar q \sigma G q \rangle \over 48}
\Big) \, .
%%%%%%%%%%%%%%%%%%%%%%%%%%%%%%%%%%%%%%%%%%%%%%%%%%%%%%%%%%%%%%%%%%%%%%%%%%%%%%%
%------------------------------\J11---------------------------------
\end{eqnarray}

\item The operator $J_{\alpha}^{1{-+}}$ couples to both $J^{PC} = 1^{-+}$ and $0^{++}$ states. The corresponding QCD sum rule equations for the quark content $\bar{q}s$ ($q = u/d$) are
\begin{eqnarray}
\nonumber
{\Pi_{1{-+}}} &=& \int^{s_0}_{m_s^2} \Bigg [
\Big(  + { m_s^2 \over 4 \pi^2 } + { m_q^2 \over 4 \pi^2 } - { m_s m_q\over 2 \pi^2 } \Big) s
-{ 9 m_s^2 m_q^2 \over 4 \pi^2}  
+ {\langle g_s^2 GG \rangle \over 32 \pi^2}
 \Bigg ] e^{-s/M^2} ds
\\ \nonumber &&
+\Big(
-{ 2 m_s m_q^2 \langle \bar s s \rangle  }
-{ 2 m_q m_s^2 \langle \bar q q \rangle  }
+{ m_q \langle g_s \bar s \sigma G s \rangle \over 2 }
+{ m_s \langle g_s \bar q \sigma G q \rangle \over 2 }
\Big) 
\\  &&
+ {16 \pi^2 \over M_B^2} \Big(  
-{ 3 \langle \bar q q \rangle m_q \langle \bar s s \rangle m_s \over 32}
+{\langle \bar q q \rangle \langle g_s \bar s \sigma G s \rangle \over 32}
+{\langle \bar s s \rangle \langle g_s \bar q \sigma G q \rangle \over 32}
\Big) \, ,
%%%%%%%%%%%%%%%%%%%%%%%%%%%%%%%%%%%%%%%%%%%%%%%%%%%%%%%%%%%%%%%%%%%%%%%%%%%%%%%
%------------------------------\J12---------------------------------
\\\nonumber
\Pi_{0{++}} &=& \int^{s_0}_{m_s^2} \Bigg [
- { 3 s^2 \over 8 \pi^2}
+ \Big(+ {3 m_s^2 \over 2 \pi^2 } + {3 m_q^2 \over 2 \pi^2 } -{3 m_q m_s \over 4 \pi^2 }\Big) s
- {9 m_s^2 m_q^2 \over2 \pi^2 } - {5 m_q \langle \bar q q \rangle \over2 } - {5 m_s \langle \bar s s \rangle \over2 }
\\ \nonumber &&
+ { m_q \langle \bar s s \rangle }
+ { m_s \langle \bar q q \rangle } - {\langle g_s^2 GG \rangle \over 48 \pi^2} \Big)
 \Bigg ] e^{-s/M^2} ds
+\Big(
+{3 m_q \langle g_s \bar s \sigma G s \rangle \over 2 \pi^2}
+{3 m_s \langle g_s \bar q \sigma G q \rangle \over 2 \pi^2}
\Big) 
\\  &&
+ {16 \pi^2 \over M_B^2} \Big(  
-{ 3 \langle \bar q q \rangle m_q \langle \bar s s \rangle m_s \over 32}
+{\langle \bar q q \rangle \langle g_s \bar s \sigma G s \rangle \over 32}
+{\langle \bar s s \rangle \langle g_s \bar q \sigma G q \rangle \over 32}
\Big) \, .
%%%%%%%%%%%%%%%%%%%%%%%%%%%%%%%%%%%%%%%%%%%%%%%%%%%%%%%%%%%%%%%%%%%%%%%%%%%%%%%
%------------------------------\J13---------------------------------
\end{eqnarray}

\item The operator $J_{\alpha}^{1{++}}$ couples to both $J^{PC} = 1^{++}$ and $0^{-+}$ states. The corresponding QCD sum rule equations for the quark content $\bar{q}s$ ($q = u/d$) are
\begin{eqnarray}
\nonumber
{\Pi_{1{++}}} &=& \int^{s_0}_{m_s^2} \Bigg [
\Big(  - { m_s^2 \over 4 \pi^2 } - { m_q^2 \over 4 \pi^2 } - { m_s m_q\over 2 \pi^2 } \Big) s
+{ 9 m_s^2 m_q^2 \over 4 \pi^2}  
- {\langle g_s^2 GG \rangle \over 32 \pi^2}
 \Bigg ] e^{-s/M^2} ds
\\ \nonumber &&
+\Big(
+{ 2 m_s m_q^2 \langle \bar s s \rangle  }
+{ 2 m_q m_s^2 \langle \bar q q \rangle  }
+{ m_q \langle g_s \bar s \sigma G s \rangle \over 2 }
+{ m_s \langle g_s \bar q \sigma G q \rangle \over 2 }
\Big) 
\\  &&
+ {16 \pi^2 \over M_B^2} \Big(  
+{ 3 \langle \bar q q \rangle m_q \langle \bar s s \rangle m_s \over 32}
+{\langle \bar q q \rangle \langle g_s \bar s \sigma G s \rangle \over 32}
+{\langle \bar s s \rangle \langle g_s \bar q \sigma G q \rangle \over 32}
\Big) \, ,
%%%%%%%%%%%%%%%%%%%%%%%%%%%%%%%%%%%%%%%%%%%%%%%%%%%%%%%%%%%%%%%%%%%%%%%%%%%%%%%
%------------------------------\J14---------------------------------
\\\nonumber
{\Pi_{0{-+}}} &=& \int^{s_0}_{m_s^2} \Bigg [
+ { 3 s^2 \over 8 \pi^2}
+ \Big(- {3 m_s^2 \over 2 \pi^2 } - {3 m_q^2 \over 2 \pi^2 } -{3 m_q m_s \over 4 \pi^2 }\Big) s
+ {9 m_s^2 m_q^2 \over2 \pi^2 } + {5 m_q \langle \bar q q \rangle \over2 } + {5 m_s \langle \bar s s \rangle \over2 }
\\ \nonumber &&
+ { m_q \langle \bar s s \rangle }
+ { m_s \langle \bar q q \rangle } + {\langle g_s^2 GG \rangle \over 48 \pi^2} \Big)
 \Bigg ] e^{-s/M^2} ds
+\Big(
+{3 m_q \langle g_s \bar s \sigma G s \rangle \over 2 \pi^2}
+{3 m_s \langle g_s \bar q \sigma G q \rangle \over 2 \pi^2}
\Big) 
\\  &&
+ {16 \pi^2 \over M_B^2} \Big(  
+{ 3 \langle \bar q q \rangle m_q \langle \bar s s \rangle m_s \over 32}
+{\langle \bar q q \rangle \langle g_s \bar s \sigma G s \rangle \over 32}
+{\langle \bar s s \rangle \langle g_s \bar q \sigma G q \rangle \over 32}
\Big) \, .
\end{eqnarray}
\end{itemize}
\end{widetext}

%
%%%%%%%%%%%%%%%%%%%%%%%%%%%%%%%%%%%%%%%%%%%%%%%%%%%%%%%%%%%%%%%%%%%%%%%%%%%%%%

%%%%%%%%%%%%%%%%%%%%%%%%%%%%%%%%%%%%%%%%%%%%%%%%%%%%%%%%%%%%%%%%%%%%%%%%%%%%%%
%

\end{document}